%% file: article.tex
\documentclass[modern]{aastex63}

\usepackage{url}
\usepackage{amsmath}
\usepackage{amssymb}
\usepackage{graphicx}
\usepackage{algorithm}
\usepackage{algorithmicx}
\usepackage{algpseudocode}

\newcommand{\abs}{\mathrm{abs}}
\newcommand{\T}[1]{T_{\mathrm{#1}}}
\renewcommand{\P}[1]{P_{\mathrm{#1}}}
\newcommand{\hatT}[1]{\hat{T}_{\mathrm{#1}}}
\newcommand{\eff}[1]{\eta_{\mathrm{#1}}}
\newcommand{\subrm}[1]{_{\mathrm{#1}}}

\newcommand{\lrparen}[1]{\left(#1\right)}
\newcommand{\lrbracket}[1]{\left[#1\right]}
\newcommand{\lrbrace}[1]{\left\{#1\right\}}
\newcommand{\uu}[2]{\underset{#1}{\underline{#2}}}

\newcommand{\dms}[3]{#1^{\circ}\, #2'\, #3''}
\newcommand{\hms}[3]{#1^{\text{h}}\, #2^{\text{m}}\, #3^{\text{s}}}

\graphicspath{{figures/}}

\accepted{July 6, 2021}
\submitjournal{AJ}
\shortauthors{Taniguchi et al.}
\shorttitle{Noise-removal method for submm spectroscopy}

\begin{document}
    \input{sections/frontmatter}
    \input{sections/introduction}
    \input{sections/reformulation}

    \input{sections/method}
    \input{sections/demonstration}
    \input{sections/discussion}
    \input{sections/conclusions}
    \input{sections/backmatter}
    \input{sections/appendix}
\end{document}

%% file: sections/frontmatter.tex
\title{A data-scientific noise-removal method for efficient submillimeter spectroscopy with single-dish telescopes}

\correspondingauthor{Akio Taniguchi}
\email{taniguchi@a.phys.nagoya-u.ac.jp}

\author[0000-0002-9695-6183]{Akio Taniguchi}
\affiliation{Nagoya University, Graduate School of Science, Division of Particle and Astrophysical Science, Furocho, Chikusa-ku, Nagoya, Aichi 464-8602, Japan}

\author[0000-0003-4807-8117]{Yoichi Tamura}
\affiliation{Nagoya University, Graduate School of Science, Division of Particle and Astrophysical Science, Furocho, Chikusa-ku, Nagoya, Aichi 464-8602, Japan}

\author[0000-0002-2462-1448]{Shiro Ikeda}
\affiliation{Institute of Statistical Mathematics, 10-3 Midori-cho, Tachikawa, Tokyo 190-8562, Japan}
\affiliation{Department of Statistical Science, The Graduate University for Advanced Studies (SOKENDAI), 10-3 Midori-cho, Tachikawa, Tokyo 190-8562, Japan}

\author[0000-0002-4124-797X]{Tatsuya Takekoshi}
\affiliation{Kitami Institute of Technology, 165 Koen-cho, Kitami, Hokkaido 090-8507, Japan}
\affiliation{Institute of Astronomy, Graduate School of Science, University of Tokyo, 2-21-1 Osawa, Mitaka, Tokyo 181-0015, Japan}

\author[0000-0002-8049-7525]{Ryohei Kawabe}
\affiliation{National Astronomical Observatory of Japan, Mitaka, Tokyo 181-8588, Japan}
\affiliation{Department of Astronomical Science, The Graduate University for Advanced Studies (SOKENDAI), 2-21-1 Osawa, Mitaka, Tokyo 181-0015, Japan}

\begin{abstract}
    For submillimeter spectroscopy with ground-based single-dish telescopes, removing noise contribution from the Earth's atmosphere and the instrument is essential.
    For this purpose, here we propose a new method based on a data-scientific approach.
    The key technique is statistical matrix decomposition that automatically separates the signals of astronomical emission lines from the drift noise components in the fast-sampled (1--10~Hz) time-series spectra obtained by a position-switching (PSW) observation.
    Because the proposed method does not apply subtraction between two sets of noisy data (i.e., on-source and off-source spectra), it improves the observation sensitivity by a factor of $\sqrt{2}$.
    It also reduces artificial signals such as baseline ripples on a spectrum, which may also help to improve the effective sensitivity.
    We demonstrate this improvement by using the spectroscopic data of emission lines toward a high-redshift galaxy observed with a 2-mm receiver on the 50-m Large Millimeter Telescope (LMT).
    Since the proposed method is carried out offline and no additional measurements are required, it offers an instant improvement on the spectra reduced so far with the conventional method.
    It also enables efficient deep spectroscopy driven by the future 50-m class large submillimeter single-dish telescopes, where fast PSW observations by mechanical antenna or mirror drive are difficult to achieve.
\end{abstract}


%% file: sections/introduction.tex
\section{Introduction}
\label{s:introduction}

Spectroscopy with large single-dish telescopes at the submillimeter wavelength is the key to understanding the dust-obscured cosmic star-formation history of the universe.
The wide-area ($> 1~\text{deg}^{2}$) and sensitive ($\sigma\sim 0.1~\text{mJy}$) spectroscopic mapping of emission lines (e.g., [C~II], [O~III]) from galaxies enables us to measure the cosmic star-formation rate density in the epoch of reionization and beyond \citep[e.g.][]{Kohno2019}.
To cover the huge three-dimensional volume of the universe (i.e., transverse area and time), large next-generation ground-based submillimeter telescopes ($D\sim50$~m) have been proposed \citep{Kawabe2016, Klaassen2019, Lou2020}.

Several new spectroscopic instruments for ``3D imagers'' have also been developed.
Wideband ($\sim 20~\text{GHz}$) spectroscopy at a high spectral resolution ($\lambda / \Delta \lambda \gtrsim 10^{5}$) becomes promising by the recent development of heterodyne receivers with instantaneous wideband \citep{Kojima2020} in radio frequency and digital spectrometers \citep{Klein2012, Iwai2017}.
As ultra-wideband ($\gtrsim 100~\text{GHz}$) spectrometers at a medium spectral resolution ($\lambda / \Delta \lambda \sim 500$), integrated superconducting spectrometers (ISSs) based on an on-chip filterbank and microwave kinetic inductance detectors (MKIDs) have been developed for blind redshift surveys \citep{Wheeler2016, Endo2019a}.
The deep spectroscopic high-redshift mapper (DESHIMA) demonstrates the detection of astronomical signals in the 345~GHz band with an instantaneous bandwidth of 45~GHz on the ASTE 10-m telescope \citep{Endo2019b} and will be upgraded to cover 220--440~GHz \citep{Endo2020}.

Alongside the new wideband instruments, observation strategies for them must be updated.
Ground-based spectroscopy at submillimeter wavelengths is strongly affected by intense molecular emission from the Earth's atmosphere (e.g., H$_{2}$O, O$_{3}$), and removing such noise emission from observed spectra is essential to obtain astronomical signals.
Calibration of the instrument (i.e., bandpass characteristics and absolute intensity scale) is also of great importance.
For these purposes, the position-switching (PSW) method is widely used for heterodyne receivers \citep{Wilson2012}, where the atmospheric emission at the sky position with an astronomical target (on-source position, hereafter) is removed through subtraction at the sky position without the target (off-source position, hereafter).
The two positions are alternatively observed generally by mechanical antenna driving at a switching interval of several seconds.
The brightness temperature of the astronomical signals corrected for the atmospheric transmission, $T_{\star}$, is obtained by the following operation:
\begin{equation}
    T_{\star}
    = \T{sys} \, \frac{\P{on} - \P{off}}{\P{off}},
    \label{eq:psw-and-cw}
\end{equation}
where $\T{sys}$ is the system noise temperature, and $\P{on}$ and $\P{off}$ are the measured powers of the on-source and off-source positions, respectively.
The standard-deviation noise level of the signals, $\sigma_{\star}$, is dependent on the frequency channel width of a spectrometer, $\Delta \nu$, and the total on-source time, $t\subrm{on}$:
\begin{equation}
    \sigma_{\star} = \frac{\sqrt{2} \, \T{sys}}{\sqrt{\Delta \nu \, t\subrm{on}}}.
    \label{eq:psw-and-cw-noise}
\end{equation}

Several observing methods have also been proposed.
The frequency switching (FSW) alternatively obtains spectra with two or more \citep{Heiles2007} different frequencies and subtracts each other.
Because it does not need to point to the off-source position, efficient on-source integration is possible in emission-line observations with moderate line widths\footnote{Maximum observable line width is limited by the maximum frequency throw (typically several tens of MHz).} ($\lesssim 100$~km~s$^{-1}$).
As an improvement of the PSW method, the wobbler switching (WSW) is used in the IRAM 30-m telescope \citep{Ungerechts2000}, where faster switching interval (typically 0.5--1~s) is achieved by wobbling the secondary mirror.

The difficulty arises, however, for these ``switching'' observations with the new instruments.
They apply subtraction of the observed noisy spectra, which results in the ``addition'' of noises to the on-source spectrum (we will refer to it as ``the direct on-off subtraction'').
This is why the factor $\sqrt{2}$ appears on the right-hand side of Equation~\ref{eq:psw-and-cw-noise}.
They also assume that the conditions of the atmosphere and the instrument (i.e., instrumental response and $\T{sys}$) should be constant during each period of switching.
In particular, the atmosphere is troublesome because its typical time-variation scale\footnote{For example, assuming a typical wind speed of $v \sim 10$~m~s$^{-1}$ and $D \sim 10\text{--}50$~m, the transverse time of a cloud across the near field of the telescope's beam is $D/v \sim 1\text{--}5$~s.} is often close to the switching interval ($10^{-1}\text{--}10^{0}$~Hz, e.g., \citet{Chapin2013}).
When the switching interval between two positions or frequencies is insufficient, drift noise contribution from the atmosphere and the instrument may cause imbalance in subtraction \citep{Schieder2001}, which results in artificial signal in the spectrum (e.g., baseline ripples).
This may even increase the effective noise level of the spectrum more than expected.
Because fast position switching by mechanical antenna or mirror drive, or large frequency throw is not realistic for wideband spectroscopy with large single-dish telescopes, current observing methods are still the bottleneck.

To overcome these issues, it is essential to develop a new method which can separate astronomical signals from drift noise components.
One of the promising solutions is to utilize a spectral correlation of the atmospheric emission:
The atmospheric spectrum depends mostly on the amount of line-of-sight water vapor, which can be expressed as a simple function.
If we continuously obtain the on-source and off-source spectra and view them as a matrix, it is well approximated with a low-rank matrix (see also Figure~\ref{fig:lowrank-sparse-decomposition}).
By contrast, deep spectroscopic surveys are expected to detect only a few bright emission lines per target source, which suggests that emission-line components should be sparse (i.e., the fraction of non-zero elements is small) in a matrix composed of on-source and off-source spectra.

In data science, sparsity-based methods have revolutionized many applications.
The innovative methods, Lasso \citep{Tibshirani1996} and compressive sensing \citep{CandesTao2006, Donoho2006} focused on the sparsity of vectors and have been successfully applied to astronomy \citep{Uemura2015, EHTcollaborationIV}.
The concept of sparsity has been generalized to matrices \citep{Candes2011}, where the sparsity of the singular values (i.e., low rank) and the component-wise sparsity have been utilized.
This extension enabled us to decompose a matrix into into low-rank and sparse components which made a variety of applications in computer vision.
We have seen some successful applications in astronomy \citep{Morii2017, Zuo2018} and our present approach also take the advantage of the effective method.

For the decomposition of a matrix into low-rank and sparse components, several computational algorithms have been proposed \citep{Candes2011, Zhou2011}.
This leads us to a possible solution:
If we can (1) obtain fast-sampled time-series spectra of on-source and off-source measurements and (2) make a matrix that can be expressed as the sum of atmospheric and emission-line components, then we would achieve continuous and noiseless estimates of the atmospheric emission even during a switching observation.

In this paper, we propose a noise-removal method for submillimeter single-dish spectroscopy based on low-rank and sparse decomposition of a matrix of a fast-sampled time-series observation.
In Section~\ref{s:reformulation}, we reformulate the PSW method and derive a time-series matrix that fulfills the conditions (1) and (2).
Section~\ref{s:proposed-method} describes algorithms for low-rank and sparse decomposition for fast-sampled PSW observations.
In Section~\ref{s:demonstration}, we apply the proposed method to real-observed fast-sampled PSW data and demonstrate that it can improve the observation sensitivity compared to the conventional direct on-off subtraction.
Finally, we discuss the advantages, limitations, and potential applications of the proposed method in Section~\ref{s:discussion}.

%% file: sections/reformulation.tex
\section{Reformulation}
\label{s:reformulation}

\begin{figure*}[t]
    \centering
    \includegraphics[width=\linewidth]{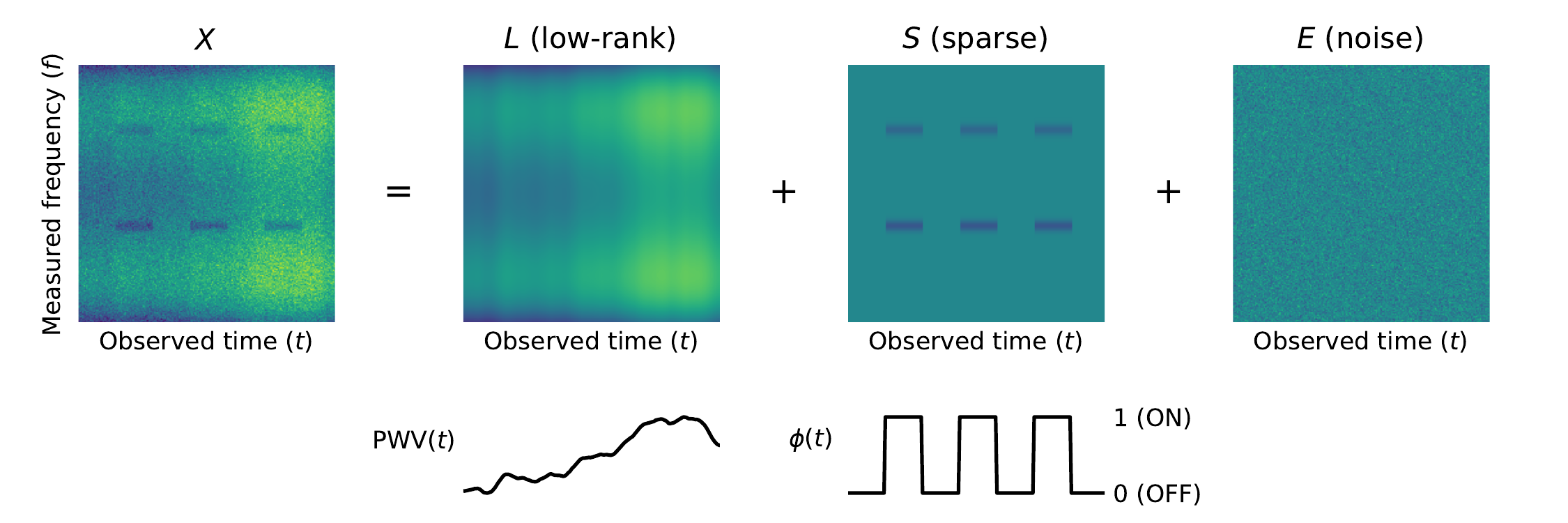}
    \caption{
        Schematic diagram of low-rank and sparse decomposition of a mock fast-sampled PSW observation.
        The top-left image represents a matrix of a dimensionless quantity, $X(f, t)$ (Equation~\ref{eq:matrix-for-decomposition}), where six rectangular features are two Gaussian-shaped emission lines observed by three on-source measurements.
        The other images represent decomposed low-rank ($L$), sparse ($S$), and noise ($E$) matrices.
        The bottom plots show the PWV and position indicator as a function of the observed time.
        Note that the values of the emission lines (non-zero values in $S$) are negative because of the definition of $X(f, t)$.
        Also note that emission lines here is set to be very bright for visualization:
        In real observations, they are often too faint and buried in the noise, the case of which will be discussed in Section~\ref{subs:godec-algorithm-for-fast-sampled-psw-observations}.
    }
    \label{fig:lowrank-sparse-decomposition}
\end{figure*}

The main idea of the proposed method is summarized in Figure~\ref{fig:lowrank-sparse-decomposition}, which schematically describes low-rank and sparse decomposition applied for a mock fast-sampled PSW observation.
We start with the general response functions (often referred to as observation equations) of a system that expresses the relation between input astronomical signals and output power spectrum of a spectrometer (Section~\ref{subs:general-response-functions-of-submillimeter-spectroscopy}).
We then reformulate the PSW method to derive an observable expressed as a product of input signals and response terms (Section~\ref{subs:reformulation-of-the-psw-method}).
Finally, we apply low-rank and sparse decomposition to the observable in a logarithmic space, derive the integrated spectrum of the mock observation, and compare it with that reduced by the direct on-off subtraction (Section~\ref{subs:signal-estimate-by-low-rank-and-sparse-matrix-decomposition}).

\subsection{General response functions of submillimeter spectroscopy}
\label{subs:general-response-functions-of-submillimeter-spectroscopy}

Let us start with the response functions of a system whose inputs are astronomical signals and outputs are the power spectrum of a spectrometer.
In the following, we use a notation in which the input signal at an observed frequency, $\nu$ (radio frequency), is measured as the output at a measured frequency, $f$ (i.e., intermediate frequency).
The output power spectrum, $P(f, t)$, is a function of the measured frequency, $f$, and time, $t$.
We assume that the response is linear, which applies to many cases of heterodyne receivers.
This means that the system output is expressed as a single gain function, $G(\nu, f)$, which applies frequency conversion and amplification using the following equation:
\begin{equation}
    P(f, t) = G(\nu, f) \, k\subrm{B} \, \lrparen{\T{in}(\nu, t) + \T{noise}(\nu)},
    \label{eq:general-response-function}
\end{equation}
where $k\subrm{B}$ is the Boltzmann constant, $\T{in}$ is the brightness temperature input to a receiver, and $\T{noise}$ is the equivalent noise temperature of the system.
We define the frequency conversion from $\nu$ to $f$ as
\begin{equation}
    \nu = f + m,
    \label{eq:frequency-conversion}
\end{equation}
where $m$ is a frequency conversion term.
In addition, $|m|$ is equivalent to the local oscillator (LO) frequency in a heterodyne receiver.
This is expressed as a function of time, $m(t)$, in frequency-modulation observations \citep{Taniguchi2020}; however, we do not consider the case hereafter (an application to this case is discussed in Section~\ref{s:discussion}).

To correct for the gain, a black body at room temperature is measured as a calibrator, which is often referred to as a hot-load measurement or chopper-wheel calibration.
Thus, there are two cases for the definition of the input temperature:
\begin{equation}
    \T{in}(\nu, t) =
    \begin{cases}
        \eff{fwd}(\nu) \, \T{sky}(\nu, t) + \lrparen{1-\eff{fwd}(\nu)} \, \T{amb} & \text{(sky measurement)}\\
        \T{room} & \text{(calibrator measurement)},
    \end{cases}
    \label{eq:input-temperature}
\end{equation}
where $\eff{fwd}$ is the forward efficiency of a telescope feed, $\T{sky}$ is the brightness temperature of signals from the sky, $\T{amb}$ is the ambient temperature at a telescope site, and $\T{room}$ is the room temperature around the receiver.
We assume that the room temperature is constant during an observation.

There are also two cases for the definition of $\T{sky}$:
\begin{equation}
    \T{sky}(\nu, t) =
    \begin{cases}
        \eff{atm}(\nu, t) \, T_{\star}(\nu, t) + \lrparen{1-\eff{atm}(\nu, t)} \, \T{atm} & \text{(on-source measurement)}\\
        \lrparen{1-\eff{atm}(\nu, t)} \, \T{atm} & \text{(off-source measurement)},
    \end{cases}
    \label{eq:sky-temperature}
\end{equation}
where $\eff{atm}$ is the line-of-sight atmospheric transmission and $\T{atm}$ is the physical temperature of the atmosphere.
In addition, $T_{\star}$ is the same as that defined in Equation~\ref{eq:psw-and-cw} and is the term of interest.
Because astronomical signals may change during an observation (e.g., on-the-fly mapping observations), it is expressed as a function of both time and frequency.

\subsection{Reformulation of the PSW method}
\label{subs:reformulation-of-the-psw-method}

We reformulate the PSW method (Equation~\ref{eq:psw-and-cw}) to derive an observable as a product of input signals and response terms such as gain and efficiency.
The keys are to unify the two cases of $\T{sky}$ (Equation~\ref{eq:sky-temperature}) into a single expression and to take the difference between the sky and calibrator measurements, rather than between on-source and off-source measurements.
Hereafter, we make the following assumptions:
\begin{align}
    \T{atm} = \T{amb} = \T{room}
    \label{eq:first-assumption}\\
    T_{\star}(\nu, t) < \T{atm}.
    \label{eq:second-assumption}
\end{align}
The former is the same as what is assumed in Equation~\ref{eq:psw-and-cw} and is used for the equation transformations in Section~\ref{subsubs:difference-between-sky-and-calibrator-measurements}.
Although we can still apply the proposed method in the case of $\T{atm} \neq \T{amb}$, this may affect the accuracy of absolute intensity scale (see also Section~\ref{subs:calibration-accuracy}).
We also assume that such physical temperatures should be estimated by another measurement, such as a thermometer or a weather monitor.
The latter is a condition that ensures the matrix decomposition described in Section~\ref{subs:signal-estimate-by-low-rank-and-sparse-matrix-decomposition}.

\subsubsection{Unification of sky measurements}
\label{subsubs:unification-of-sky-measurements}

We start by introducing a position indicator, $\phi(t)$, which discriminates between on-source and off-source times during an observation:
\begin{equation}
    \phi(t) =
    \begin{cases}
        1 & \text{(when pointing to on-source position)}\\
        0 & \text{(when pointing to off-source position)}.
    \end{cases}
    \label{eq:position-indicator}
\end{equation}
Practically stated, it should be derived from an antenna log of the observation.
Using the position indicator, we define unified astronomical signals, $\T{ast}$, which express both on-source and off-source measurements at the same time:
\begin{equation}
    \T{ast}(\nu, t) =
    \begin{cases}
        T_{\star}(\nu, t) & (\phi(t) = 1)\\
        0 & (\phi(t) = 0),
    \end{cases}
    \label{eq:signal-temperature}
\end{equation}
and $\T{sky}$ is then expressed as
\begin{equation}
    \T{sky}(\nu, t)
    = \eff{atm}(\nu, t) \, \T{ast}(\nu, t) + \lrparen{1-\eff{atm}(\nu, t)} \, \T{atm}.
    \label{eq:unified-sky-temperature}
\end{equation}

\subsubsection{Difference between sky and calibrator measurements}
\label{subsubs:difference-between-sky-and-calibrator-measurements}

We take the difference between the measurements of the sky and a calibrator to express a new observable for the following noise-removal method.
As expressed in Equation~\ref{eq:psw-and-cw}, the PSW method takes the difference between on-source and off-source measurements.
By contrast, the proposed sky-calibrator difference is expressed as a product of input signals and response terms, including the gain and efficiency.
Using the equations above, we express the output power spectra measured for the sky and a calibrator as follows:
\begin{align}
    \P{sky}(f, t)
    &= G(\nu, f) \, k\subrm{B} \, \lrbracket{\eff{fwd}(\nu) \, \T{sky}(\nu, t) + \lrparen{1-\eff{fwd}(\nu)} \, \T{amb} + \T{noise}(\nu)}
    \label{eq:output-of-signal}\\
    \P{cal}(f, t)
    &= G(\nu, f) \, k\subrm{B} \, \lrbracket{\T{room} + \T{noise}(\nu)}.
    \label{eq:output-of-calibration}
\end{align}
The sky-calibrator difference, $dP(f, t)$, is then expressed as follows:
\begin{align}
    dP(f, t)
    & \equiv \P{sky}(f, t) - \P{cal}(f, t)
    \label{eq:difference-between-outputs}\\
    & = G(\nu, f) \, k\subrm{B} \, \lrbracket{\eff{fwd}(\nu) \, \lrparen{\T{sky}(\nu, t) - \T{amb}} + \uu{= \, 0}{\T{amb} - \T{room}}}\notag\\
    & = G(\nu, f) \, k\subrm{B} \, \eff{fwd}(\nu) \, \lrbracket{\eff{atm}(\nu, t) \, \lrparen{\T{ast}(\nu, t) - \T{atm}} + \uu{= \, 0}{\T{atm} - \T{amb}}}\notag\\
    & = G(\nu, f) \, k\subrm{B} \, \eff{fwd}(\nu) \, \eff{atm}(\nu, t) \, \lrparen{\T{ast}(\nu, t) - \T{atm}}.
    \label{eq:difference-between-outputs-with-assumptions}
\end{align}

Note that we ignore the cosmic microwave background (CMB) emission in the definition of $\T{ast}$ (Equation~\ref{eq:signal-temperature}).
If it is not negligible (e.g., observations at lower frequencies close to millimeter wavelength), $\T{atm}$ in Equation~\ref{eq:difference-between-outputs-with-assumptions} should be replaced with $\T{atm} - \T{cmb}(\nu)$, where $\T{cmb}(\nu)$ is the brightness temperature of the CMB.

\subsection{Signal estimate by low-rank and sparse matrix decomposition}
\label{subs:signal-estimate-by-low-rank-and-sparse-matrix-decomposition}

From Equation~\ref{eq:difference-between-outputs-with-assumptions}, $dP(f, t)$ is a product of the signal term $(\T{ast} - \T{atm})$ and the response terms $(G k\subrm{B} \eff{fwd} \eff{atm})$.
Here we introduce a dimensionless quantity, $X(f, t)$, as follows:
\begin{align}
    X(f, t)
    & \equiv \ln \lrparen{-\frac{dP(f, t)}{k\subrm{B} \, \T{atm}}}\notag\\
    & = \ln \lrbrace{G(\nu, f) \, \eff{fwd}(\nu) \, \eff{atm}(\nu, t)} + \ln \lrparen{1-\frac{\T{ast}(\nu, t)}{\T{atm}}}.
    \label{eq:matrix-for-decomposition}
\end{align}
This is a sum of two terms.
The first term varies as the atmosphere fluctuates over time and always has non-zero values.
As demonstrated in many submillimeter measurements \citep{Dempsey2013, Cortes2016}, the logarithm of $\eff{atm}(\nu, t)$ (i.e., the atmospheric opacity) is typically expressed as a linear function of line-of-sight precipitable water vapor, $\mathrm{PWV}(t)$:
\begin{equation}
    \ln \, \eff{atm}(\nu, t) \simeq a(\nu) \, \mathrm{PWV}(t) + b(\nu),
\end{equation}
where $a(\nu)$ and $b(\nu)$ are coefficients depending on $\nu$.
When $X(f, t)$ is discretely sampled in the time and frequency domains, it suggests that the first term of Equation~\ref{eq:matrix-for-decomposition} should be a low-rank matrix.
By contrast, the second term is a sparse matrix because its elements only become non-zero when a telescope is pointing to the on-source position (i.e., $\phi(t) = 1$) and emission lines exist (i.e., $T_{\star}(\nu, t) \neq 0$).
If the sparseness is ensured during an observation, $X(f, t)$ can be decomposed into low-rank ($L$) and sparse ($S$) components:
\begin{equation}
    X = L + S + E,
    \label{eq:lowrank-sparse-decomposition}
\end{equation}
where $E$ is a noise term\footnote{Here, $E$ appears because the noise terms (e.g., $\T{noise}(\nu)$) are not constant but stochastic, which may follow the normal distribution. The sky-calibrator difference (Equations~\ref{eq:difference-between-outputs}--\ref{eq:matrix-for-decomposition}) actually has an additional term related to $E$, but it is not explicitly described for simplicity.}.
$L$ and $S$ correspond to the following:
\begin{align}
    L(\nu, f, t) &\simeq \ln \lrparen{G(\nu, f) \, \eff{fwd}(\nu) \, \eff{atm}(\nu, t)}
    \label{eq:estimate-of-lowrank}\\
    S(\nu, t) &\simeq \ln \lrparen{1-\frac{\T{ast}(\nu, t)}{\T{atm}}}.
    \label{eq:estimate-of-sparse}
\end{align}

Although $\T{ast}(\nu, t)$ is derived from $S(\nu, t)$, it is a noiseless estimate:
If one prefers astronomical signals with errors as the final product, it is reasonable that the estimate of astronomical signals should be composed of both $S$ and $E$.
Finally, we define the estimate of astronomical signals, $\hatT{ast}$, using the low-rank component:
\begin{equation}
    \hatT{ast}(\nu, t) = \T{atm} \, \lrbracket{1-\exp\lrparen{X(f, t) - L(\nu, f, t)}}.
    \label{eq:estimate-of-signal}
\end{equation}
For PSW observations, the integrated spectrum, $\bar{T}_{\star}(\nu)$, is derived as
\begin{equation}
    \bar{T}_{\star}(\nu) = \frac{\int_{0}^{t\subrm{obs}} \, \hatT{ast}(\nu, t) \, \phi(t) \, dt}{\int_{0}^{t\subrm{obs}} \, \phi(t) \, dt},
    \label{eq:integrated-spectrum}
\end{equation}
where $t\subrm{obs}$ is the total observation time of both on and off-source positions.

\begin{figure}[t]
    \centering
    \includegraphics[width=\linewidth]{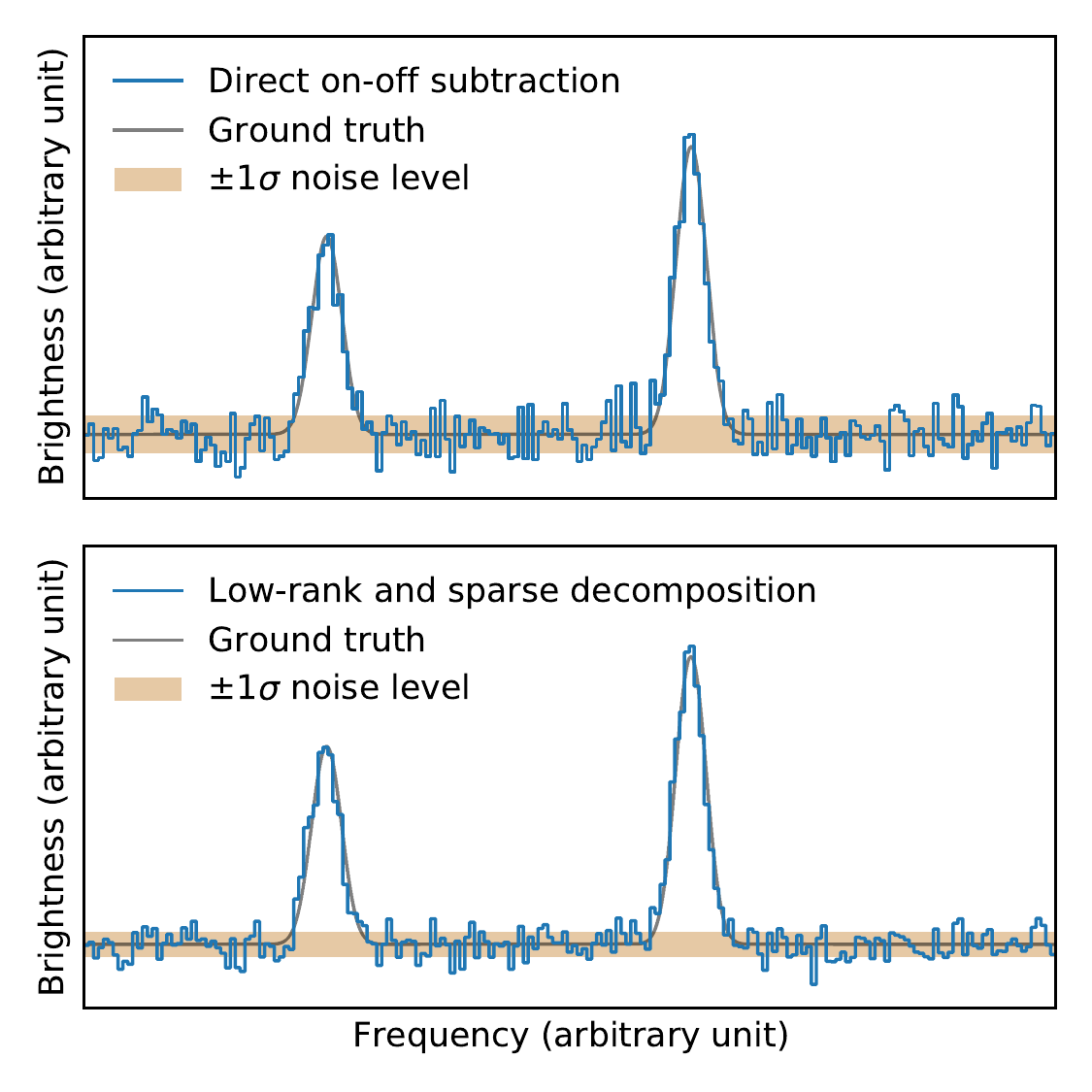}
    \caption{
        Integrated spectra of the mock observation (the same data in Figure~\ref{fig:lowrank-sparse-decomposition}) reduced by (top) the conventional direct on-off subtraction using Equation~\ref{eq:psw-and-cw} and (bottom) low-rank and sparse decomposition using Equation~\ref{eq:estimate-of-signal}.
        Gray lines are the ground truth of the astronomical signals.
        Dim orange spans indicate standard deviations of line-free channels of the spectra.
        Note that we independently create $L$, $S$, and $E$ for Figure~\ref{fig:lowrank-sparse-decomposition} and the spectra are made from them.
        This means that the spectrum of the bottom panel is not estimated from $X$ using the GoDec algorithm shown in Section~\ref{s:proposed-method}.
    }
    \label{fig:spectra-mock-observation}
\end{figure}

Figure~\ref{fig:spectra-mock-observation} shows the integrated spectra of the mock observation (the same data in Figure~\ref{fig:lowrank-sparse-decomposition}) reduced by the proposed decomposition (Equation~\ref{eq:estimate-of-signal}) and the conventional direct on-off subtraction (Equation~\ref{eq:psw-and-cw}).
In this case, the noise level is improved by a factor of 1.44 ($\simeq \sqrt{2}$) while keeping the spectral shape of the astronomical signals.

The signal estimation works better when the astronomical signals of interest are sparse.
The assurance of sparseness depends on the observing mode (e.g., fraction of on-source time) and spatial and spectral distribution of an astronomical target (e.g., number of channels the emission line enters and/or compactness of the target in the case of a mapping observation).
In Sections~\ref{s:proposed-method} and \ref{s:demonstration}, we demonstrate the case of fast-sampled PSW observations toward high-redshift galaxies where only a single emission line exists within a spectral band.
Other cases such as frequency-modulation observations are discussed in Section~\ref{s:discussion}.

%% file: sections/method.tex
\section{Proposed method}
\label{s:proposed-method}

We describe the proposed method to achieve low-rank and sparse decomposition in this section.
The method is based on the GoDec algorithm \citep{Zhou2011}.
We modify the sparse-identification step of the algorithm for the fast-sampled PSW observations.
After introducing the mathematical notation in Section~\ref{subs:mathematical-expression-of-time-series-spectra}, we describe the algorithm for estimating astronomical signals from fast-sampled PSW observations in Section~\ref{subs:godec-algorithm-for-fast-sampled-psw-observations}.

\subsection{Mathematical expression of time-series spectra}
\label{subs:mathematical-expression-of-time-series-spectra}

The time-series spectra of a fast-sampled PSW observation is represented as a discretely sampled matrix.
For example, the dimensionless quantity, $X(f, t)$, is expressed as
\begin{equation}
    X(f, t) \rightarrow x_{ij} \quad (1 \leq i \leq N\subrm{freq}, 1 \leq j \leq N\subrm{time}),
    \label{eq:discrete-sample}
\end{equation}
where $N\subrm{freq}$ and $N\subrm{time}$ are the number of samples of the frequency and time domains, respectively.
We use a bold upper-case letter to denote a matrix:
\begin{equation}
    \boldsymbol{X} = (x_{ij}) \in \mathbb{R}^{N\subrm{freq} \times N\subrm{time}}.
\end{equation}
The $j$-th column vector of a matrix corresponding to a sample spectrum is shown as a bold lower-case letter such as $\boldsymbol{x}_{j}$.
The position indicator, $\phi(t)$, is also time-sampled and expressed as an $N\subrm{time}$-dimension column vector, $\boldsymbol{\phi}$.
For other notations, $\boldsymbol{0}$ and $\boldsymbol{1}$ are matrices with all elements being zero and one, respectively.
An element-wise product of two matrices (Hadamard product) is $\boldsymbol{X} \circ \boldsymbol{Y} \equiv (x_{ij} y_{ij})$.
$\mathrm{abs}(\cdot)$ and $\exp(\cdot)$ are element-wise absolute-value and natural-exponential functions, respectively.
The Frobenius norm of a matrix (the extension of a vector norm to a matrix) is expressed as $||\boldsymbol{X}||_{F} \equiv (\sum_{i} \sum_{j} x_{ij}^{2})^{1/2}$.

Using the notations, low-rank and sparse decomposition (Equation~\ref{eq:lowrank-sparse-decomposition}) becomes
\begin{equation}
    \boldsymbol{X} = \boldsymbol{L} + \boldsymbol{S} + \boldsymbol{E}.
\end{equation}
The estimates of astronomical signals and an integrated spectrum (Equations~\ref{eq:estimate-of-signal}--\ref{eq:integrated-spectrum}), which we finally aim to derive, are expressed as follows:
\begin{align}
    \hat{\boldsymbol{T}}\subrm{ast} &= \T{atm} \lrparen{\boldsymbol{1} - \exp(\boldsymbol{X} - \boldsymbol{L})},\\
    \bar{\boldsymbol{t}}_{\star} &= (\boldsymbol{\phi}^{T} \boldsymbol{\phi})^{-1} \, \hat{\boldsymbol{T}}\subrm{ast} \, \boldsymbol{\phi}.
\end{align}

\subsection{GoDec algorithm for fast-sampled PSW observations}
\label{subs:godec-algorithm-for-fast-sampled-psw-observations}

\begin{figure}[t]
    \centering
    \begin{algorithm}[H]
        \caption{GoDec algorithm}
        \label{algo:godec-algorithm}
        \begin{algorithmic}[1]
            \Require $\boldsymbol{X} \in \mathbb{R}^{N\subrm{freq} \times N\subrm{time}}, \, r, \, k, \, \varepsilon$
            \Ensure $\boldsymbol{L}, \, \boldsymbol{S} \in \mathbb{R}^{N\subrm{freq} \times N\subrm{time}}$
            \State $n = 0, \, \boldsymbol{L}_{0} = \boldsymbol{X}, \, \boldsymbol{S}_{0} = \boldsymbol{0}$
            \While{$||\boldsymbol{X} - \boldsymbol{L}_{n} - \boldsymbol{S}_{n}||_{F}^{2} \, / \, ||\boldsymbol{X}||_{F}^{2} > \varepsilon$}
                \State $\boldsymbol{L}_{n+1} = \sum_{j=1}^{r} \lambda_{jj} \, \boldsymbol{u}_{j} \, \boldsymbol{v}_{j}^{T}$ s.t. $\boldsymbol{U} \boldsymbol{\Lambda} \boldsymbol{V}^{T} = \mathrm{SVD}(\boldsymbol{X} - \boldsymbol{S}_{n})$
                \State $\boldsymbol{\Omega}_{n+1} = \mathrm{SparseID}(\abs(\boldsymbol{X} - \boldsymbol{L}_{n+1}); k)$
                \State $\boldsymbol{S}_{n+1} = \boldsymbol{\Omega}_{n+1} \circ (\boldsymbol{X} - \boldsymbol{L}_{n+1})$
                \State $n := n + 1$
            \EndWhile
            \State \Return $\boldsymbol{L}_{n}$, $\boldsymbol{S}_{n}$
        \end{algorithmic}
    \end{algorithm}
\end{figure}

We describe the GoDec algorithm and modify a part of it for estimating the astronomical signals of fast-sampled PSW observations.
It is an iterative algorithm in which $\boldsymbol{L}$ and $\boldsymbol{S}$ are alternatively estimated, i.e., it assigns the low-rank approximation of $\boldsymbol{X} - \boldsymbol{S}$ to $\boldsymbol{L}$ and the sparse approximation of $\boldsymbol{X} - \boldsymbol{L}$ to $\boldsymbol{X}$.
Algorithm~\ref{algo:godec-algorithm} shows the GoDec algorithm (the notation is modified from \citet{Zhou2011} to fit the present paper), where $r$ is the rank of a low-rank matrix, $k$ is the number of non-zero elements of a sparse matrix, and $\varepsilon$ is a threshold for the convergence of the algorithm.
In addition, $\mathrm{SVD}(\cdot)$ conducts singular value decomposition\footnote{Singular value decomposition (SVD) is one of the matrix decomposition methods, where it decompose a matrix into (left and right) singular vectors. Because the first $r$ vectors represent $r$ most dominant components in the matrix, it is sometimes used for low-rank approximation or dimensionality reduction of a matrix.}, where $\boldsymbol{\Lambda}$ is a diagonal matrix of singular values, and $\boldsymbol{U}$ and $\boldsymbol{V}$ are left and right singular vectors, respectively.
$\textsc{SparseID}(\cdot\,; k)$ is a sparse-identification step to compute an index matrix, $\boldsymbol{\Omega} = (\omega_{ij} \in \{0, 1\})$, whose elements only become unity if the corresponding elements of the input matrix are the first $k$-largest elements.

\begin{figure*}[t]
    \centering
    \includegraphics[width=\linewidth]{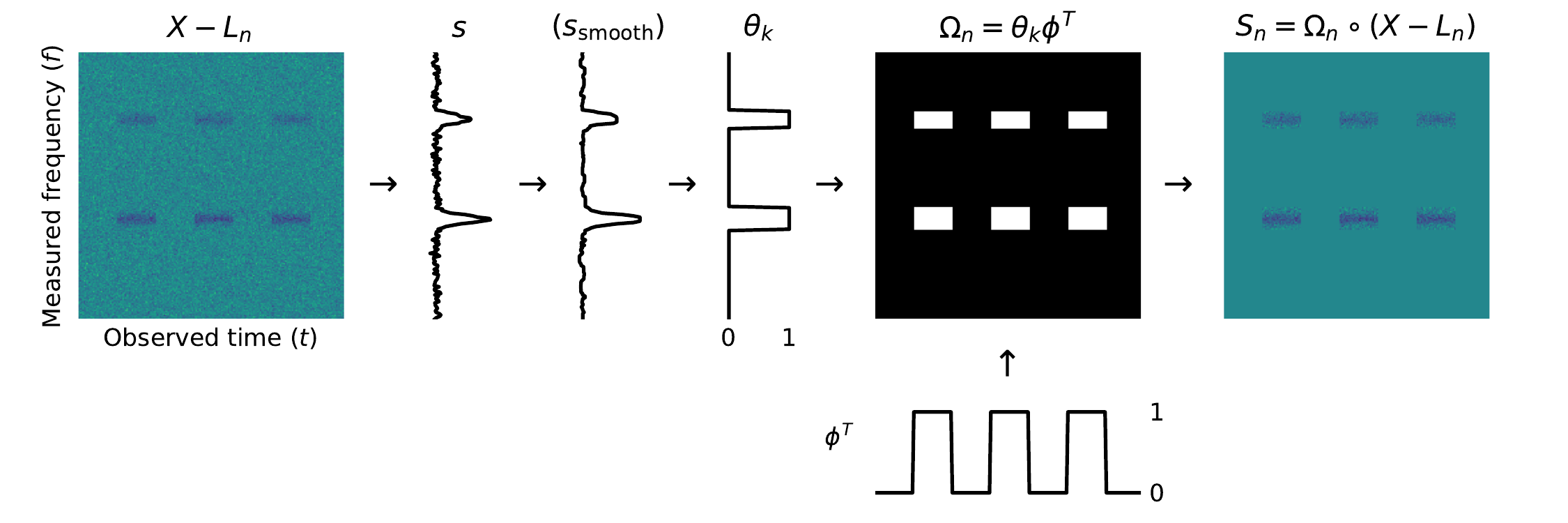}
    \caption{
        Schematic diagram of sparse identification for fast-sampled PSW observations (Algorithm~\ref{algo:sparseid-for-fast-sampled-psw-observations}).
        The top-left image represents a low-rank-subtracted matrix that should contain both astronomical signals and noise.
        In addition, $\boldsymbol{s}$ is a time-integrated on-source spectrum, and $\boldsymbol{s}\subrm{smooth}$ is a spectrum smoothed by a low-pass-like filter (optional step).
        Sparse identification (finding the first $k$ largest elements) is applied in the spectrum ($\boldsymbol{\theta}_{k}$), and the index matrix, $\boldsymbol{\Omega}$, is computed as an outer product of $\boldsymbol{\theta}_{k}$ and $\boldsymbol{\phi}$.
    }
    \label{fig:sparseid-for-fast-sampled-psw-observations}
\end{figure*}

\begin{figure}[t]
    \centering
    \begin{algorithm}[H]
        \caption{Sparse identification for fast-sampled PSW observations}
        \label{algo:sparseid-for-fast-sampled-psw-observations}
        \begin{algorithmic}[1]
            \Require $\boldsymbol{Y} \in \mathbb{R}^{N\subrm{freq} \times N\subrm{time}}$, $\boldsymbol{\phi} \in \{0, 1\}^{N\subrm{time}}$, $k$
            \Ensure $\boldsymbol{\Omega} \in \{0, 1\}^{N\subrm{freq} \times N\subrm{time}}$
            \Function{SparseID@PSW}{$\boldsymbol{Y}$; $\boldsymbol{\phi}$, $k$}
                \State $\boldsymbol{s} = (\boldsymbol{\phi}^{T} \boldsymbol{\phi})^{-1} \, \boldsymbol{Y} \boldsymbol{\phi}$
                \State $\boldsymbol{\theta}_{k} = \mathrm{SparseID}(\abs(\boldsymbol{s}); k)$
                \State $\boldsymbol{\Omega} = \boldsymbol{\theta}_{k} \, \boldsymbol{\phi}^{T}$
                \State \Return $\boldsymbol{\Omega}$
            \EndFunction
        \end{algorithmic}
    \end{algorithm}
\end{figure}

As the result of the GoDec algorithm, we expect the sparse matrix to extract the astronomical signals (Figure~\ref{fig:lowrank-sparse-decomposition}).
In real observations, however, astronomical signals are often much weaker than the noise level in a spectrum at a sampling interval of $\sim$1~s:
In high-redshift observations with a 50-m class telescope, the typical peak intensity of an emission line is $\sim 1$~mK, and the signal-to-noise ratio (SNR) in a 1-s integrated spectrum is $\mathrm{SNR} \sim 10^{-2}$.
This indicates that the original sparse-identification step should be modified for the current purpose.

Herein, we propose a custom sparse-identification step, $\textsc{SparseID@PSW}(\cdot\,; \boldsymbol{\phi}, k)$, which employs information on the position indicator and the time-integrated spectrum, and replace $\textsc{SparseID}$ with it.
Figure~\ref{fig:sparseid-for-fast-sampled-psw-observations} shows a schematic diagram of the step, where $\boldsymbol{s}$ is a column vector of the time-integrated on-source spectrum and $\boldsymbol{\theta}_{k}$ is a column vector of the index spectrum to indicate the first $k$-largest elements.
The index matrix, $\boldsymbol{\Omega}$, is computed as the outer product of $\boldsymbol{\theta}_{k}$ and $\boldsymbol{\phi}$ (i.e., the intersection between on-source times and signal-detected frequencies).
Algorithm~\ref{algo:sparseid-for-fast-sampled-psw-observations} describes the step.
Note that the number of non-zero elements in $\boldsymbol{S}$ is not $k$ but $k (\boldsymbol{\phi}^{T}\boldsymbol{\phi})$.

As we will see in Section~\ref{s:demonstration}, the custom sparse-identification step enables us to identify astronomical signals even when $\mathrm{SNR} \ll 1$.
The parameters ($r$, $k$) should be set a priori (i.e., hyperparameters) or optimized, for example, through a cross validation.
The rank of the low-rank matrix, $r$, depends on the conditions of the atmosphere, the observed bandwidth, and the observed frequency.
\citet{Taniguchi2020} demonstrates that $r \simeq 5$ is typically sufficient to remove correlated noise components in a frequency-modulation observation using the Nobeyama 45-m telescope, which ensures that $r$ is much less than $\min(N\subrm{freq}, N\subrm{time}) \sim 10^{3}$.
The number of non-zero elements in a time-integrated spectrum, $k$, depends on the number of channels at which emission lines are detected.
In the case of a fast-sampled PSW observation toward a high-redshift emission line with a typical full width at half maximum (FWHM) of 1000~km~s$^{-1}$, using a 4-GHz bandwidth in the 345~GHz band, the fraction of non-zero elements in $\boldsymbol{\Omega}$ (and thus $\boldsymbol{S}$) is $\sim$15\%, which assures the sparseness.

As illustrated in Figure~\ref{fig:sparseid-for-fast-sampled-psw-observations}, smoothing a time-integrated spectrum before sparse identification is promising in the case of observations toward emission lines that have broader and fainter features (e.g., an outflow).
In this case, the window length of a smoothing filer, $w$, is another parameter, which depends on the noise level of the spectrum or the spectral shape of the emission lines.
We demonstrate the data reduction of actual fast-sampled PSW observations with spectral smoothing in Section~\ref{s:demonstration}.

%% file: sections/demonstration.tex
\section{Demonstration}
\label{s:demonstration}

We show an application of the proposed algorithms (Section~\ref{s:proposed-method}) to fast-sampled PSW observations toward a high-redshift galaxy.
We start by describing the conditions of the observations and data reduction using the conventional method (direct on-off subtraction and linear-baseline fitting) and the proposed method (Section~\ref{subs:data-description-and-reduction}).
We then show the integrated spectra of the reduced data and demonstrate the improvement of the observation sensitivity achieved by the proposed method (Section~\ref{subs:improvement-of-observation-sensitivity}).
Finally, we investigate the noise characteristics of the reduced data to show the effect of low-rank noise removal (Section~\ref{subs:characteristics-of-noise-on-time-series-spectra}).

\subsection{Data description and reduction}
\label{subs:data-description-and-reduction}

\begin{table*}[t]
    \centering
    \caption{Observation logs of the target taken by the B4R on the 50-m LMT}
    \label{tab:observation-logs-of-the-target-taken-by-b4r-on-lmt}
    \begin{tabular}{p{0.5\linewidth}p{0.2\linewidth}p{0.2\linewidth}}
        \hline\hline
        ~ & CO~($J = 4\text{--}3$) & CO~($J = 5\text{--}4$)\\
        \hline
        Target name & \multicolumn{2}{l}{PJ020941.3}\\
        Target coordinates (J2000) & \multicolumn{2}{l}{$\alpha=\hms{02}{09}{41.3}$, $\delta=\dms{+00}{15}{59}$}\\
        Off-source relative coordinates (horizontal) & \multicolumn{2}{l}{$d\text{Az}=\dms{+00}{01}{00}$, $d\text{El}=\dms{00}{00}{00}$}\\
        LMT project ID & \multicolumn{2}{l}{2019S1B4RCommissioning}\\
        LMT observation ID (hot load) & 86889 & 86895\\
        LMT observation ID (science target) & 86890 & 86896\\
        \hline
        Observation date & 2019 Nov 26 & 2019 Nov 26\\
        Observation start time (UTC) & 05:11:06 & 05:54:25\\
        Opacity at 220~GHz & 0.17 & 0.16\\
        System noise temperature (K) & 106 & 120\\
        \hline
        B4R first-LO frequency (GHz) & 137.0 & 155.3\\
        XFFTS frequency range (GHz) & 128.9--131.4 & 160.9--163.4\\
        256-ch binned channel width (GHz) & 0.02 (40~km~s$^{-1}$) & 0.02 (36~km~s$^{-1}$)\\
        XFFTS data-dumping rate (sample~s$^{-1}$) & 1.0 & 1.0\\
        \hline
        Integration time of hot load (s) & 10.0 & 10.0\\
        Integration time at on-source position (s) & 300 (10~s $\times$ 30) & 300 (10~s $\times$ 30)\\
        Integration time at off-source position (s) & 300 (10~s $\times$ 30) & 300 (10~s $\times$ 30)\\
        Transition time between two positions (s) & 5--6 & 5--6\\
        \hline
    \end{tabular}
\end{table*}

We use time-series spectra of fast-sampled PSW observations for an extremely luminous submillimeter galaxy PJ020941.3, which was originally found through submillimeter-to-far-infrared continuum imaging surveys of \textit{Planck} (\textit{Planck} Catalog of Compact Sources; \citet{Planck2014}) and \textit{Herschel} (\textit{Herschel} Stripe 82 Survey; \citet{Viero2014}).
Its spectroscopic redshift was determined to be $z\subrm{spec} = 2.5534 \pm 0.0002$ \citep{Harrington2016} through a 3-mm observation of the redshifted CO~($J = 3\text{--}2$) line ($\nu\subrm{obs} = 97.314$~GHz) using the Redshift Search Receiver (RSR; \citet{Erickson2007}) installed on the Large Millimeter Telescope (LMT; \citet{Schloerb2008}).

Our observations targeted on two redshifted emission lines of CO~($J = 4\text{--}3$) ($\nu\subrm{obs} = 129.739$~GHz) and CO~($J = 5\text{--}4$) ($\nu\subrm{obs} = 162.165$~GHz) using a 2-mm sideband-separating heterodyne receiver installed on the 50-m LMT (B4R\footnote{\url{http://lmtgtm.org/b4r/}}; Kawabe et al. in prep) connected to the XFFTS spectrometer \citep{Klein2012}.
As summarized in Table~\ref{tab:observation-logs-of-the-target-taken-by-b4r-on-lmt}, we carried out the fast-sampled PSW observations in late 2019 as part of a commissioning campaign of the B4R.
For each observation, we took time-series spectra at a sampling rate of 1~Hz using two-sideband and dual-polarization settings.
On-source and off-source spectra were alternatively observed for 10~s, and the sequence was repeated 30 times, which resulted in a total on-source time of 300~s.
A 10-s measurement of a hot load preceded the observation.
As a result, an observation yielded a matrix of continuous 600-s time-series spectra\footnote{During the observation, we had a certain amount of antenna transition time between the two positions in which the spectrometer did not obtain data. This means that the spectra are not perfectly continuous and the system may have drifted.} whose dimensions are $N\subrm{time} = 600$ and $N\subrm{freq} = 2^{15}$ (the number of channels of XFFTS).

After the observations, we carried out data reduction as follows:
For common settings, we binned 256 frequency channels of a matrix to improve the signal-to-noise ratios ($N\subrm{freq} = 128$ after binning), which resulted in a velocity resolution of $\sim$36--40~km~s$^{-1}$ (see Table~\ref{tab:observation-logs-of-the-target-taken-by-b4r-on-lmt}).
We used a physical temperature of the atmosphere of $\T{atm}=273.0$~K.
In the case of a conventional data reduction, we first time-integrated each 10-s observation of hot-load, on-source, and off-source measurements, and then applied chopper-wheel calibration to them (i.e., Equation~\ref{eq:psw-and-cw}).
This resulted in 30 spectra of calibrated astronomical signals.
Linear baseline subtraction was applied to each spectrum.
We finally time-integrated the spectra to obtain the final spectrum of the target.

In the case of the proposed data reduction, we integrated the 10-s hot-load measurement to obtain $\P{cal}$.
We then obtained a dimensionless matrix, $\boldsymbol{X}$, according to Equations~\ref{eq:difference-between-outputs}--\ref{eq:matrix-for-decomposition} and applied Algorithm~\ref{algo:godec-algorithm} to it using the custom sparse-identification step (Algorithm~\ref{algo:sparseid-for-fast-sampled-psw-observations}).
As illustrated in Figure~\ref{fig:sparseid-for-fast-sampled-psw-observations}, we used a median filter to obtain a smoothed spectrum before sparse identification, $\boldsymbol{s}\subrm{smooth} = \mathrm{filter}(\boldsymbol{s}; w)$, where $w$ is a parameter of the filter window length.
We used the parameters $(r, k, w) = (5, 25, 5)$ for the CO~(4--3) observation and $(r, k, w) = (7, 25, 7)$ for CO~(5--4).
The difference between the parameters assumed for CO~(4--3) and CO~(5--4) is attributed to a higher system noise temperature and a narrower velocity width per channel in the latter case.
Finally, we estimated the time-series spectra of astronomical signals from $\boldsymbol{L}$ and $\boldsymbol{S}$ according to Equation~\ref{eq:estimate-of-signal} and time-integrated them to obtain an integrated spectrum of the target.

\subsection{Improvement of observation sensitivity}
\label{subs:improvement-of-observation-sensitivity}

\begin{figure*}[t]
    \centering
    \includegraphics[width=\linewidth]{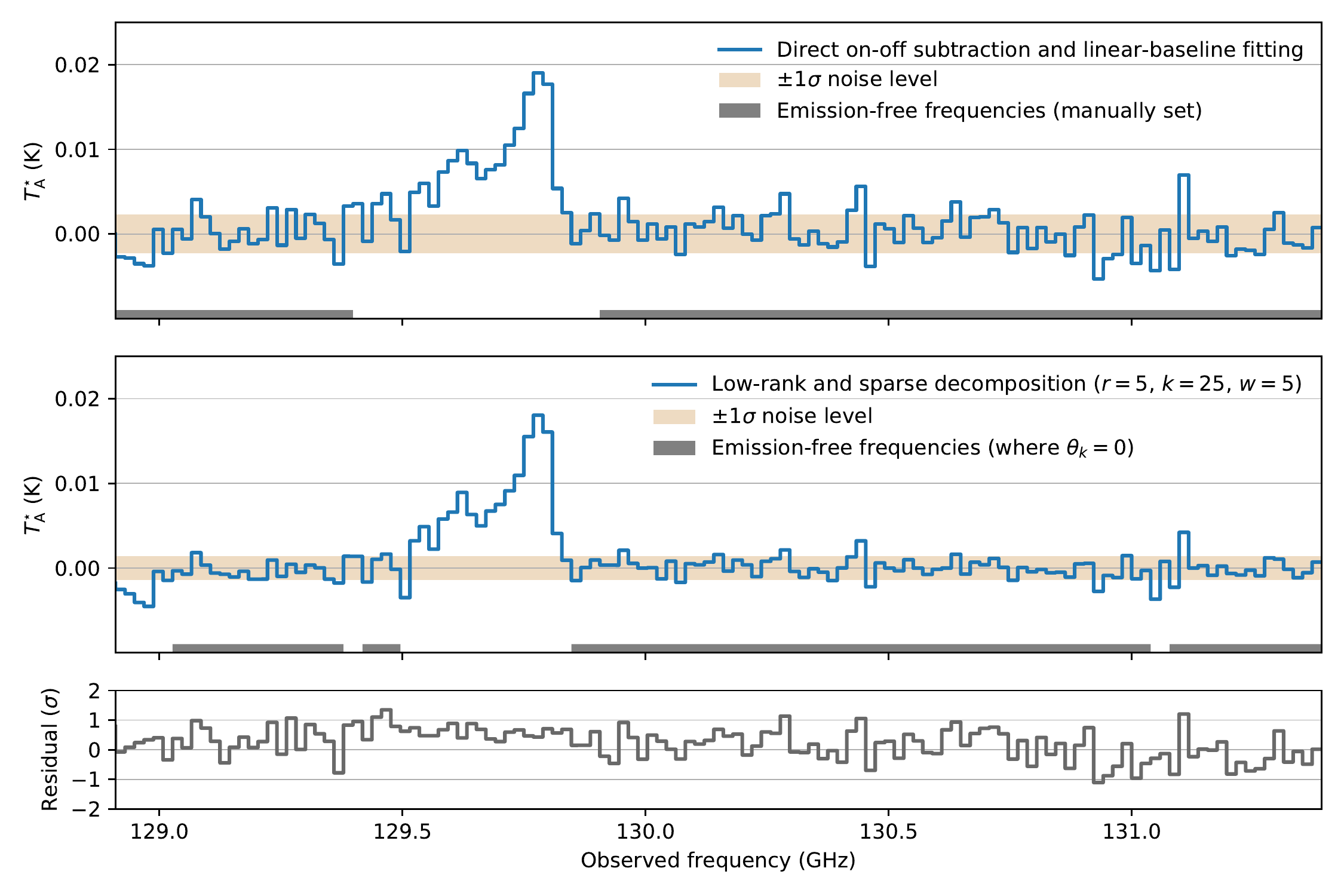}
    \caption{
        The integrated spectra of the redshifted CO~(4--3) emission line of PJ020941.3 reduced by the conventional (top) and proposed (middle) methods.
        Dim orange spans indicate the achieved standard deviations of emission-free channels of the spectra (Table~\ref{tab:noise-levels-of-the-CO43-spectra-reduced-by-the-different-methods} left column).
        The unit of the vertical axis is $\T{A}^{\star}$, which corresponds to astronomical signals corrected for atmospheric attenuation.
        Other effects (e.g., main beam and/or beam dilution) are not corrected.
        In a conventional case, we excluded a frequency range of 129.4--129.9~GHz in a linear-baseline fitting.
        Emission-free frequencies are indicated as gray strips in the plot.
        In the proposed case, the parameters for the Algorithms~\ref{algo:godec-algorithm}--\ref{algo:sparseid-for-fast-sampled-psw-observations} are $(r, k, w) = (5, 25, 5)$, where $w$ is the window length of the median filter before sparse identification.
        Estimated emission-free frequencies (i.e., where $\theta_{k} = 0$) are indicated as gray strips in the plot.
        Other parameters are listed in Table~\ref{tab:observation-logs-of-the-target-taken-by-b4r-on-lmt}.
        The bottom panel shows the difference spectrum between the two methods divided by the achieved standard deviation of the conventional method.
    }
    \label{fig:spectra-PJ020941.3-CO43}
\end{figure*}

We show the integrated spectra of the redshifted CO~(4--3) emission line of PJ020941.3 in Figure~\ref{fig:spectra-PJ020941.3-CO43} reduced by the conventional and proposed methods (those of CO~(5--4) are shown in Figure~\ref{fig:spectra-PJ020941.3-CO54}).
We detected the emission line at an observed frequency of 129.7~GHz in both methods.
The spectral shapes and intensities of the emission lines reduced by both methods are consistent with each other.
The full width at zero intensity (FWZI) of each line is $\sim$800~km~s$^{-1}$, which is consistent with that of CO~(3--2) obtained using RSR \citep{Harrington2016}.

The spectrum reduced by the proposed method in Figure~\ref{fig:spectra-PJ020941.3-CO43} shows that the noise level is improved compared to that reduced by the conventional method.
Because the proposed method does not apply on-off subtraction on noisy spectra, the factor $\sqrt{2}$ in Equation~\ref{eq:psw-and-cw-noise} is expected to decrease if the noise is ideally white.
We summarize the achieved and expected noise levels of both cases in Table~\ref{tab:noise-levels-of-the-CO43-spectra-reduced-by-the-different-methods} (those of CO~(5--4) are shown in Table~\ref{tab:noise-levels-of-the-CO54-spectra-reduced-by-the-different-methods}).
The achieved standard deviation of emission-free channels in the proposed case is lower than that of the conventional case by a factor of 1.67.
This also means that a required noise level can be achieved in a 1.67$^{2}$-times shorter observation\footnote{In the case of no overhead time. In an actual observation, constant overhead such as calibrator measurements may increase the total observation time.} with the proposed method.

The factor of 1.67 is slightly better than the expected improvement of $\sqrt{2}$.
This extra improvement may be attributed to the subtraction of low-rank characteristics of atmospheric fluctuation and/or time variation of the instrumental response, which causes a baseline fluctuation in an integrated spectrum.
As shown in Table~\ref{tab:noise-levels-of-the-CO43-spectra-reduced-by-the-different-methods}, the expected noise level inferred from $\T{sys}$ is close to the achieved value in the proposed case.
By contrast, the achieved value in the conventional case is 1.2-times larger than the expected value, which suggests the existence of baseline fluctuations.

\begin{table*}[t]
    \centering
    \caption{Noise levels of the CO~(4--3) spectra reduced by different methods}
    \label{tab:noise-levels-of-the-CO43-spectra-reduced-by-the-different-methods}
    \begin{tabular}{p{0.26\linewidth}p{0.32\linewidth}p{0.32\linewidth}}
        \hline\hline
        ~ & Achieved standard deviation of emission-free channels & Expected noise level inferred from Equation~\ref{eq:psw-and-cw-noise}\\
        ~ & (mK) & (mK)\\
        \hline
        Conventional method & 2.30 & 1.92\\
        Proposed method & 1.38 & 1.35 ($=1.92/\sqrt{2}$)\\
        \hline
    \end{tabular}
\end{table*}

\subsection{Characteristics of noise on time-series spectra}
\label{subs:characteristics-of-noise-on-time-series-spectra}

\begin{figure*}[t]
    \centering
    \includegraphics[width=\linewidth]{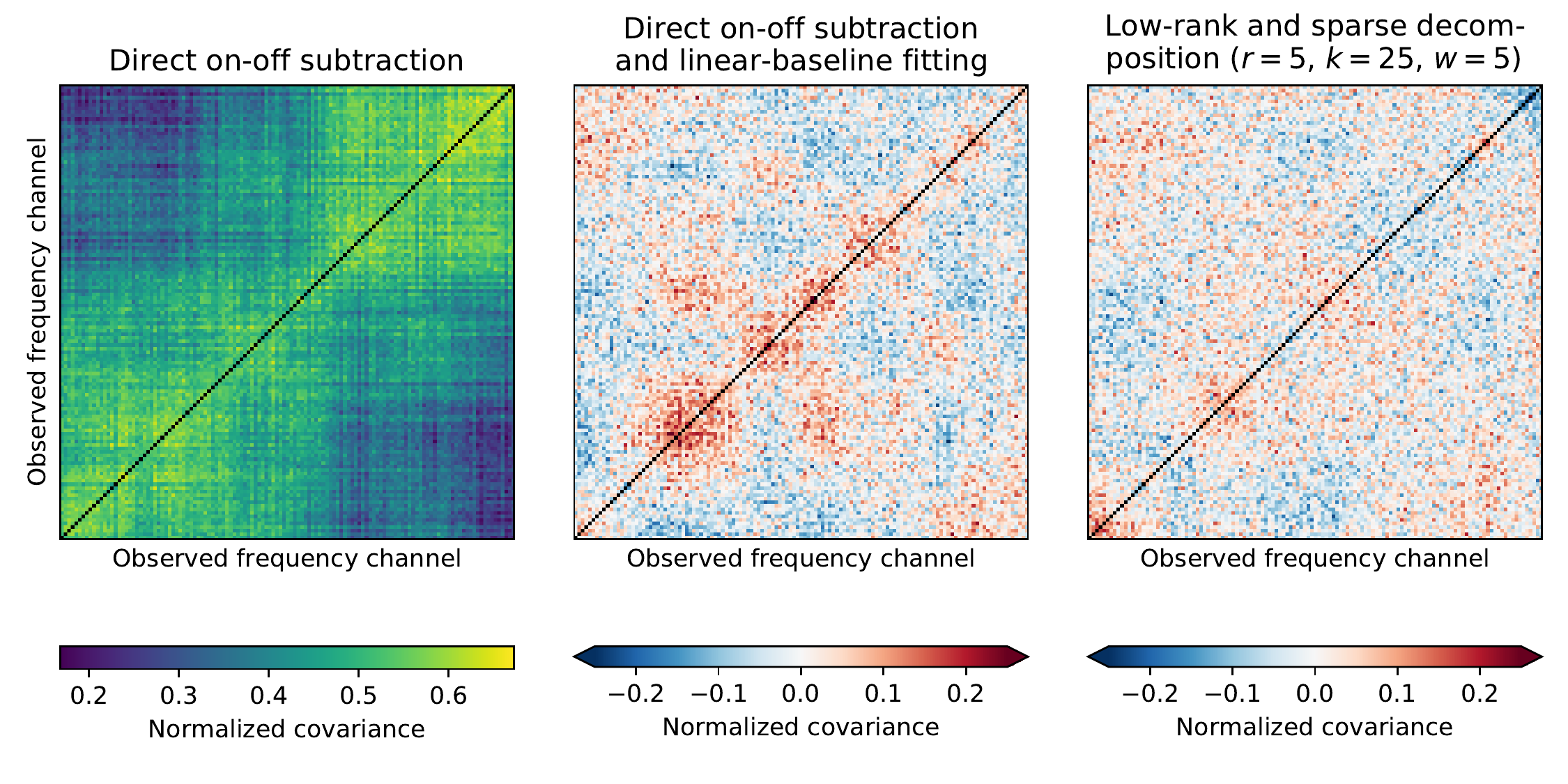}
    \caption{
        Covariance matrices of reduced time-series spectra of the redshifted CO~(4--3) observation (note that they are normalized such that the diagonal elements are unity).
        Left: reduced only through a direct on-off subtraction.
        Center: reduced using a direct on-off subtraction and a linear-baseline fitting (the conventional method).
        Right: reduced through the GoDec algorithm for fast-sampled PSW observations (the proposed method).
        The parameters for these methods are listed in Figure~\ref{fig:spectra-PJ020941.3-CO43}.
    }
    \label{fig:covariance-PJ020941.3-CO43}
\end{figure*}

\begin{figure}[t]
    \centering
    \includegraphics[width=\linewidth]{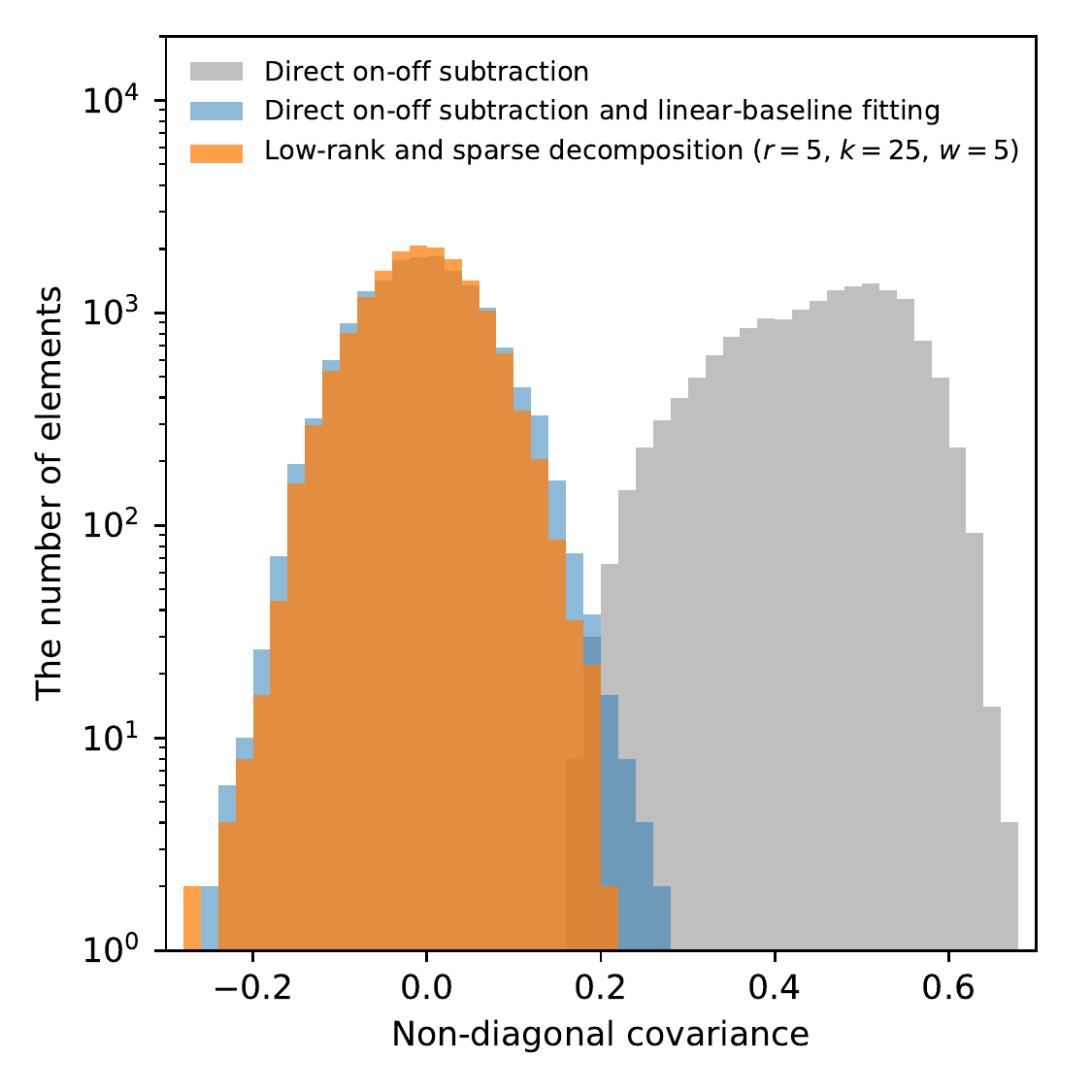}
    \caption{Histogram of non-diagonal elements of the covariance matrices in Figure~\ref{fig:covariance-PJ020941.3-CO43}.}
    \label{fig:covariance-histogram-PJ020941.3-CO43}
\end{figure}

\begin{figure}[t]
    \centering
    \includegraphics[width=\linewidth]{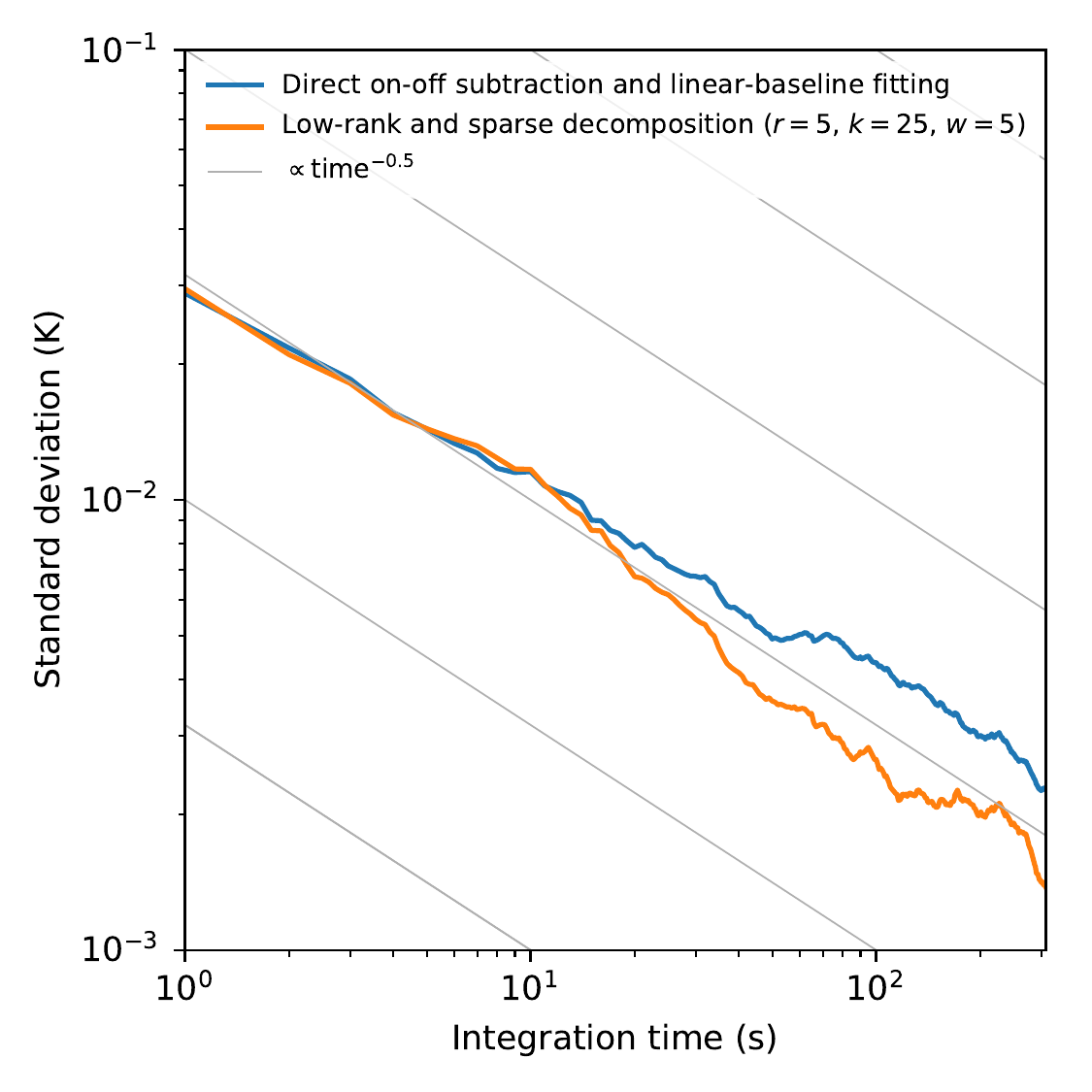}
    \caption{
        Integration time versus achieved standard deviation of emission-free channels in the time-series spectra of the redshifted CO~(4--3) observation reduced through the conventional (blue line) and proposed (orange line) methods.
        To obtain time-series spectra in the former case, we first time-integrated each 10-s off-source measurement and then subtract it from each 1-s on-source measurement.
        Gray sloped lines are proportional to the square root of the integration time (the standard deviation decreases in parallel in the case of white noise).
    }
    \label{fig:time-vs-noise-PJ020941.3-CO43}
\end{figure}

We further investigate the characteristics of noise on the reduced time-series spectra.
Figures~\ref{fig:covariance-PJ020941.3-CO43} and \ref{fig:covariance-histogram-PJ020941.3-CO43} show the covariance matrices of time-series spectra of the CO~(4--3) observation reduced by different methods and their histograms, respectively (those of CO~(5--4) are shown in Figures~\ref{fig:covariance-PJ020941.3-CO54} and \ref{fig:covariance-histogram-PJ020941.3-CO54}).
While the conventional case (center) indicates that non-diagonal covariance components (i.e., common mode noises) remain in the integrated spectrum, most of them are subtracted using the proposed method (mean absolute covariance of $<5$\%).
This improvement can also be seen in the time-versus-noise plot in Figure~\ref{fig:time-vs-noise-PJ020941.3-CO43} (that of CO~(5--4) is shown in Figure~\ref{fig:time-vs-noise-PJ020941.3-CO54}).
Although the achieved standard deviations of both cases are consistent at an integration time less than the PSW interval ($\sim$10~s), the standard deviation of the conventional case becomes worse than the expected value at a longer integration time.
This suggests the existence of atmospheric fluctuations and/or time variation of an instrumental response of slower than 0.1~Hz.

%% file: sections/discussion.tex

\section{Discussion}
\label{s:discussion}

We discuss the advantages, limitations, and possible applications of the proposed reformulation and algorithms in submillimeter spectroscopy.

\subsection{Observation sensitivity and efficiency}
\label{subs:observation-sensitivity-and-efficiency}

One of the major advantages of the proposed method is the improvement in the observation sensitivity through a post-process:
If fast-sampled time-series spectra are available, the observation sensitivity can be improved by at least a factor of $\sqrt{2}$ even if they are obtained in a past observation.
Because many single-dish telescopes already have the capability of on-the-fly mapping observations that obtain time-series spectra at $>1$~Hz, the fast-sampled PSW mode should be feasible.
Thus, the proposed method can immediately improve their observation sensitivities.

The proposed method may also improve the observation efficiency (the fraction of on-source time over the total observation time).
Because it does not require off-source measurements at a fixed sky position, the antenna transition time between the on and off-source positions can be partially used for the measurements, which was not available before.
This means that the integration time at the off-source position can be reduced, while keeping the total off-source (sky) measurement time.

As another advantage, the proposed method can improve the baseline stability because the GoDec algorithm estimates common modes (singular vectors) in a matrix of time-series spectra.
In addition to an atmospheric fluctuation, periodic baseline ripples caused by standing waves in an instrument not only worsen the observation sensitivity but they also make it difficult to distinguish emission lines from such artifacts.
The proposed method may also be a powerful baseline fitting tool where fitting functions can be automatically determined from the data themselves.

As we demonstrated using the B4R implemented on the 50-m LMT, the observation sensitivities and baseline stabilities are improved by the proposed method.
Although the targets are limited to luminous high-redshift galaxies with shorter observation times ($\sim10$~min), we show that it can be applied for single emission line observations.
In a future study, the proposed method needs to be verified for observations of longer integration times ($> 1\text{--}10 $~h) to establish it for deep spectroscopic science cases, as introduced in Section~\ref{s:introduction}.

\subsection{Calibration accuracy}
\label{subs:calibration-accuracy}

The proposed method does not improve the accuracy of absolute intensity scales compared to the conventional method because it continues to use a single hot-load calibration.
The major assumptions of the method (Equation~\ref{eq:first-assumption}) are sometimes unrealistic:
For example, $\T{atm}$ is expected to be less than $\T{amb}$ under a good weather condition.
We can still apply the method and expect the improvement, however, it may cause a systematic error in $\hatT{ast}$ in the same way as the conventional method.
This is because an additional term, $G \, k\subrm{B} \, \eff{fwd} \, \lrparen{\T{atm} - \T{amb}}$, appears in the sky-calibrator difference ($dP$; Equation~\ref{eq:difference-between-outputs-with-assumptions}), which is difficult to assume or estimate.

The systematic error can be reduced with better calibration.
For example, one can directly derive the following quantity without the assumptions of Equation~\ref{eq:first-assumption} using two-load calibration (i.e., measurements of hot and cold loads):
\begin{equation}
    \frac{\T{hot} - \T{cold}}{\T{atm}} \cdot \frac{\P{hot} - \P{sky}}{\P{hot} - \P{cold}}
    = \eff{fwd} \, \eff{atm} \lrparen{1 - \frac{\T{ast}}{\T{atm}}}
    - \eff{fwd} \, \lrparen{1 - \frac{\T{amb}}{\T{atm}}},
    \label{eq:two-load-calibration}
\end{equation}
where $\T{hot}$ and $\T{cold}$ are the temperature of the hot and cold loads, respectively.
$\P{hot}$ and $\P{cold}$ are given by Equation~\ref{eq:output-of-calibration} whose $\T{room}$ is replaced with $\T{hot}$ and $\T{cold}$, respectively.
Unlike the additional term above, the second term in Equation~\ref{eq:two-load-calibration} is close to zero in many cases.
This means that the logarithm of Equation~\ref{eq:two-load-calibration} is the sum of low-rank and sparse components and thus can be used as $X$.
In the case of a more accurate calibration, where $\T{sky}(\nu, t)$ can be directly derived (using sky-dip measurements, for example), the proposed algorithms can be applied to $\T{sky}(\nu, t) - \T{atm}$ instead of $dP(f, t)$.
This strategy would be useful if an instrument has a non-linear response to the signal (e.g., DESHIMA).

\subsection{Observation targets}
\label{sub:observation-targets}

Like other correlated noise-removal methods, the proposed method has certain limitations in terms of the spectral distribution of the emission.
Because it estimates and removes common-mode spectra, it cannot be used for continuum observations where the emission uniformly enters all spectral channels.
It also assumes the sparseness of the signal in a matrix.
As we demonstrate in Section~\ref{s:demonstration}, the signal of $\sim$15\% sparseness can be estimated properly.
This suggests that observations of a single emission line (e.g., blind redshift surveys of distant galaxies) should be promising for most broadband spectrometers.
It would be challenging, however, in the cases of multiple emission lines that occupy a substantial fraction of the spectrometer bandwidth (e.g., line surveys of nearby galaxies).
In such cases, it would still be useful if the on-source fraction was less than 50\% (e.g., mapping observations of compact sources).
The feasibility of the proposed method for different target types should be further investigated using a large dataset of either past or future observations.

\subsection{Potential applications}
\label{subs:potential-applications}

The proposed method has potential applications to many types of observations and/or instruments other than PSW observations by heterodyne receivers.
It can be applied to observations with DESHIMA and other integrated superconducting spectrometers (ISSs) where the other efficient noise-removal method based on frequency modulation \citep{Taniguchi2020} cannot be used.
Moreover, because it does not depend on spatial scan patterns of antennas, it can be applied to on-the-fly mapping observations of compact objects.
\citet{Taniguchi2020} also demonstrate that the correlated noise-removal method effectively reduces scanning effects (artificial stripes seen in an image along the scan direction) of such observations.
As mapping observations become crucial to investigate the ``3D property'' of galaxies, the proposed method with 3D imagers would be expected to accelerate the survey speed of the investigation.
An initial application of the proposed method to DESHIMA will be demonstrated in a future study.

%% file: sections/conclusions.tex
\section{Conclusions}
\label{s:conclusions}

In this paper, we propose a new noise-removal method based on low-rank and sparse decomposition to achieve continuous and noiseless estimates of the atmospheric emission for submillimeter single-dish spectroscopy.
The conclusions are as follows:

\begin{itemize}
    \item
        We propose a new noise-removal method that achieves the continuous and noiseless estimate of atmospheric emission by obtaining fast-sampled time-series spectra of on and off-source measurements in a PSW observation and applying low-rank and sparse decomposition to them (Section~\ref{s:introduction}).
    \item
        We show that the time-series astronomical spectra of a single-dish telescope with a hot-load calibration can be expressed as a sum of low-rank atmospheric emission and sparse astronomical signals by reformulating the PSW method (Section~\ref{s:reformulation}).
    \item
        We show that GoDec, one of the low-rank and sparse decomposition algorithms, can be applied to a matrix of time-series spectra with a custom sparse-identification step for fast-sampled PSW observations (Section~\ref{s:proposed-method}).
    \item
        We demonstrate that the proposed method improves the observation sensitivity by a factor of 1.67 using the data of fast-sampled PSW observations obtained by the B4R on the 50-m LMT (Section~\ref{s:demonstration}).
        We find that the improvement is better than the expected value of $\sqrt{2}$, which suggests that the proposed method can also reduce the baseline fluctuation of an integrated spectrum.
    \item
        We discuss the advantages, limitations, and potential applications of the proposed method (Section~\ref{s:discussion}), and propose the application of it to future ultra-wideband spectrometers such as DESHIMA.
	    By contrast, application to long-integrated observations has yet to be investigated, and will be demonstrated in future studies.
\end{itemize}

%% file: sections/backmatter.tex
\acknowledgments

We thank the anonymous referee for fruitful comments.
This work is supported by KAKENHI (Nos.~15H02073, 17H06130, and 20H01951).
This paper makes use of data taken by the Large Millimeter Telescope Alfonso Serrano (LMT) in Mexico.
The LMT project is a joint effort of the Instituto Nacional de Astr\'{o}fisica, \'{O}ptica, y
Electr\'{o}nica (INAOE) and the University of Massachusetts at Amherst (UMASS).
We also appreciate the support of the technical staff and the support scientists of the LMT during the commissioning campaign of the B4R.

\facility{
    LMT (B4R)
}

\software{
    Astropy \citep{Astropy2013,Astropy2018},
    NumPy \citep{Harris2020},
    matplotlib \citep{Hunter2007},
    scikit-learn \citep{Pedregosa2011},
    SciPy \citep{Virtanen2020},
    pandas \citep{McKinney2010,McKinney2011},
    xarray \citep{Hoyer2017},
}

\bibliography{references}
\bibliographystyle{aasjournal}

%% file: sections/appendix.tex
\appendix

\section{Results and analyses of the CO~(5--4) observation}
\label{s:results-and-analyses-of-the-co-(5--4)-observation}

\begin{figure*}[h]
    \centering
    \includegraphics[width=\linewidth]{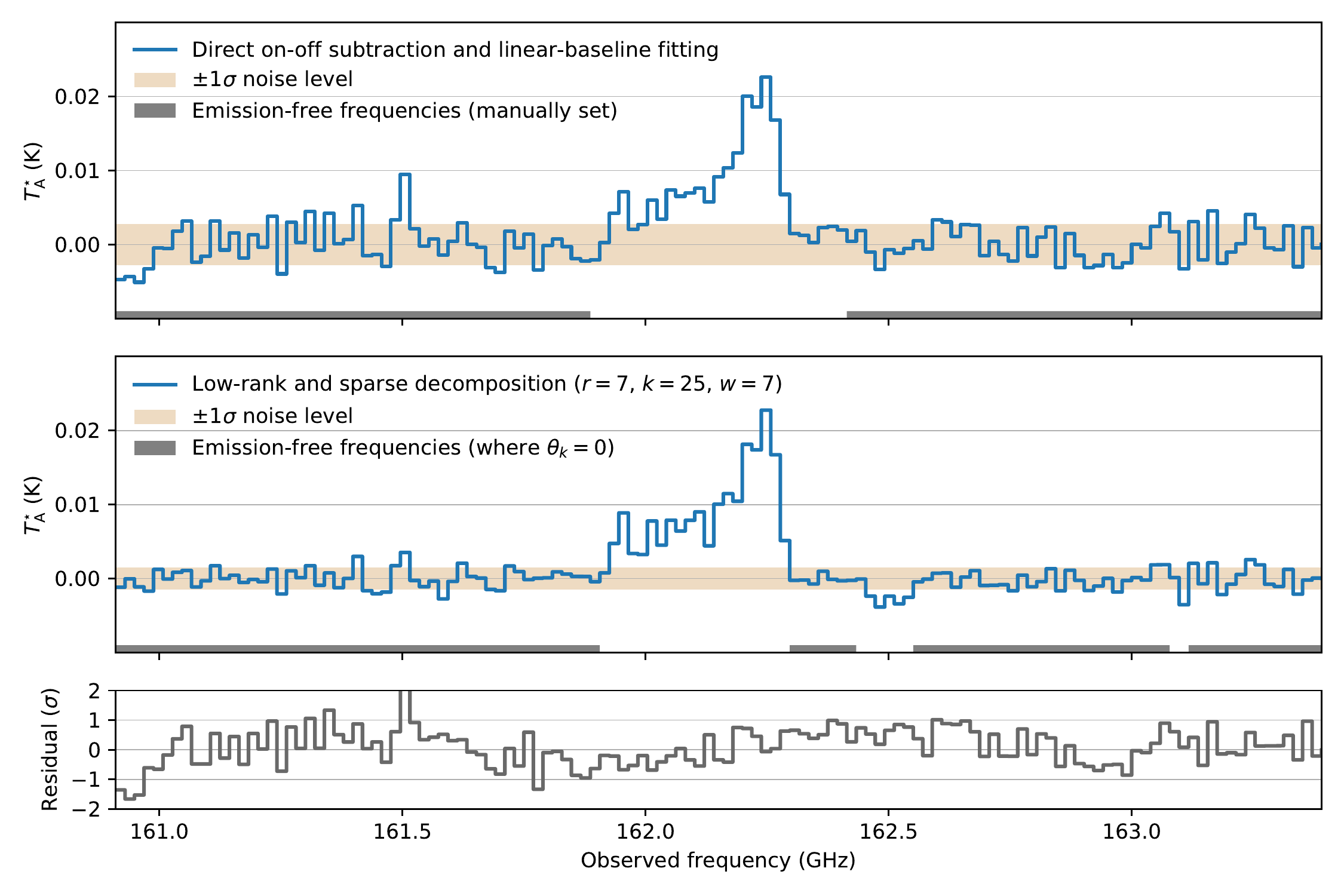}
    \caption{
        The integrated spectra of the redshifted CO~(5--4) emission line of PJ020941.3 reduced by the conventional (top) and proposed (middle) methods.
        Dim orange spans indicate the achieved standard deviations of emission-free channels of the spectra (Table~\ref{tab:noise-levels-of-the-CO54-spectra-reduced-by-the-different-methods} left column).
        The unit of the vertical axis is $\T{A}^{\star}$, which corresponds to astronomical signals corrected for atmospheric attenuation.
        Other effects (e.g., main beam and/or beam dilution) are not corrected.
        In the conventional case, we excluded a frequency range of 161.9--162.4~GHz in a linear-baseline fitting.
        Emission-free frequencies are indicated as gray strips in the plot.
        In the proposed case, the parameters for the Algorithms~\ref{algo:godec-algorithm}--\ref{algo:sparseid-for-fast-sampled-psw-observations} are $(r, k, w) = (7, 25, 7)$, where $w$ is the window length of the median filter before sparse identification.
        Estimated emission-free frequencies (i.e., where $\theta_{k} = 0$) are indicated as gray strips in the plot.
        Other parameters are listed in Table~\ref{tab:observation-logs-of-the-target-taken-by-b4r-on-lmt}.
        The bottom panel shows the difference spectrum between the two methods divided by the achieved standard deviation of the conventional method.
    }
    \label{fig:spectra-PJ020941.3-CO54}
\end{figure*}

\begin{figure*}[h]
    \centering
    \includegraphics[width=\linewidth]{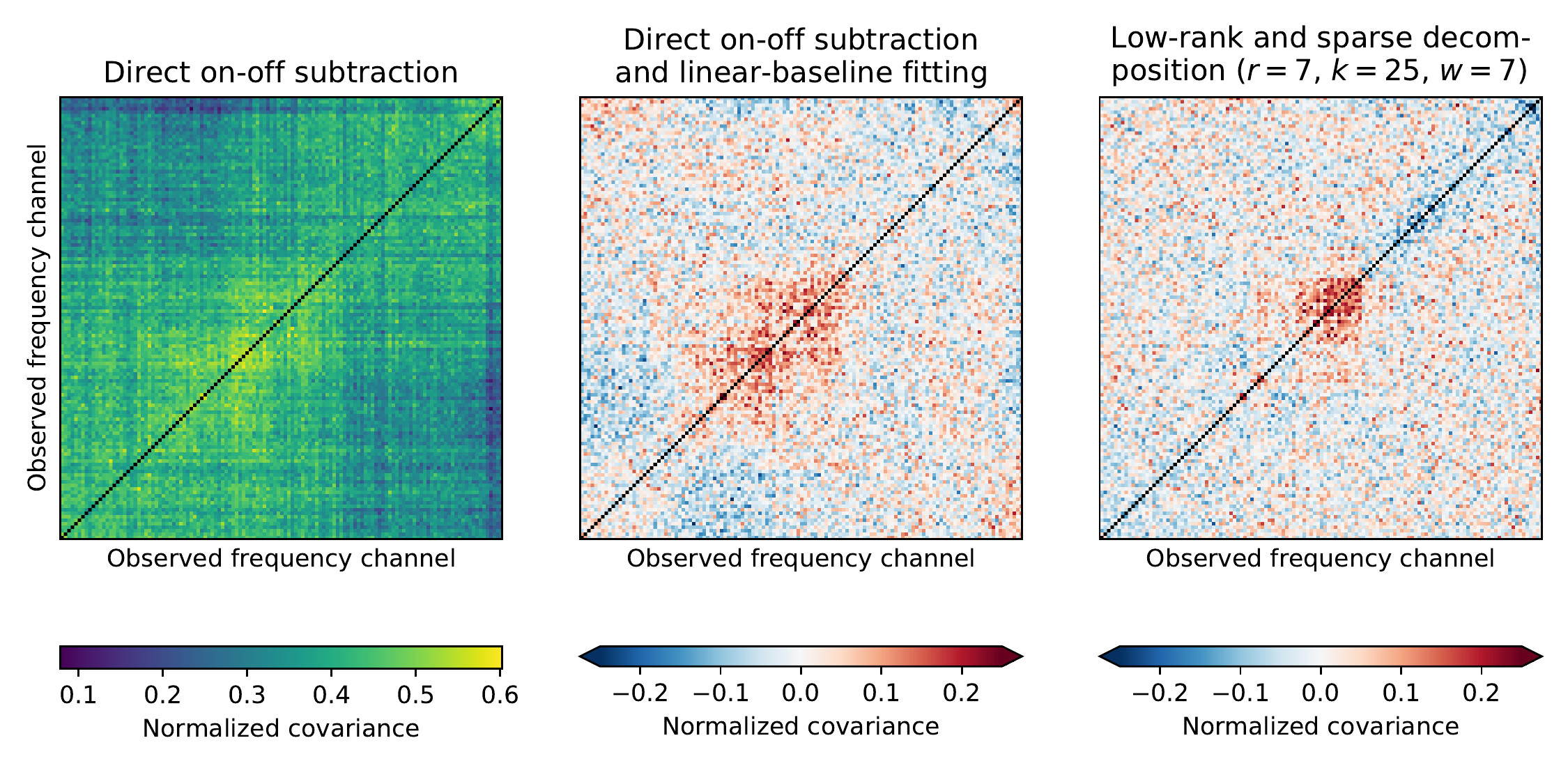}
    \caption{
        Covariance matrices of reduced time-series spectra of the redshifted CO~(5--4) observation (note that they are normalized such that the diagonal elements are unity).
        Left: reduced using only a direct on-off subtraction.
        Center: reduced through a direct on-off subtraction and a linear-baseline fitting (the conventional method).
        Right: reduced using the GoDec algorithm for fast-sampled PSW observations (the proposed method).
        The parameters for these methods are listed in Figure~\ref{fig:spectra-PJ020941.3-CO54}.
    }
    \label{fig:covariance-PJ020941.3-CO54}
\end{figure*}

\begin{figure}[h]
    \centering
    \includegraphics[width=\linewidth]{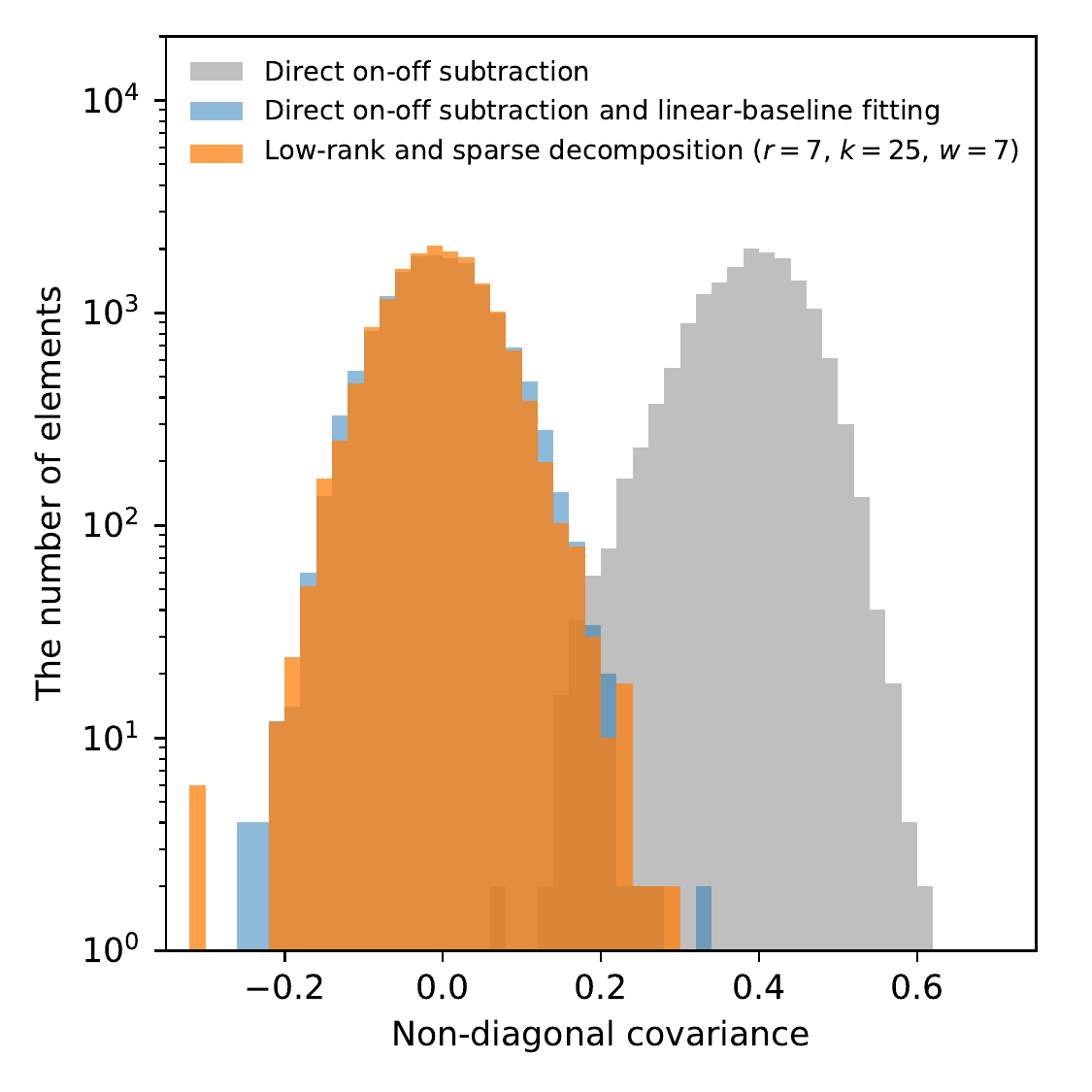}
    \caption{Histogram of non-diagonal elements of the covariance matrices in Figure~\ref{fig:covariance-PJ020941.3-CO54}.}
    \label{fig:covariance-histogram-PJ020941.3-CO54}
\end{figure}

\begin{figure}[h]
    \centering
    \includegraphics[width=\linewidth]{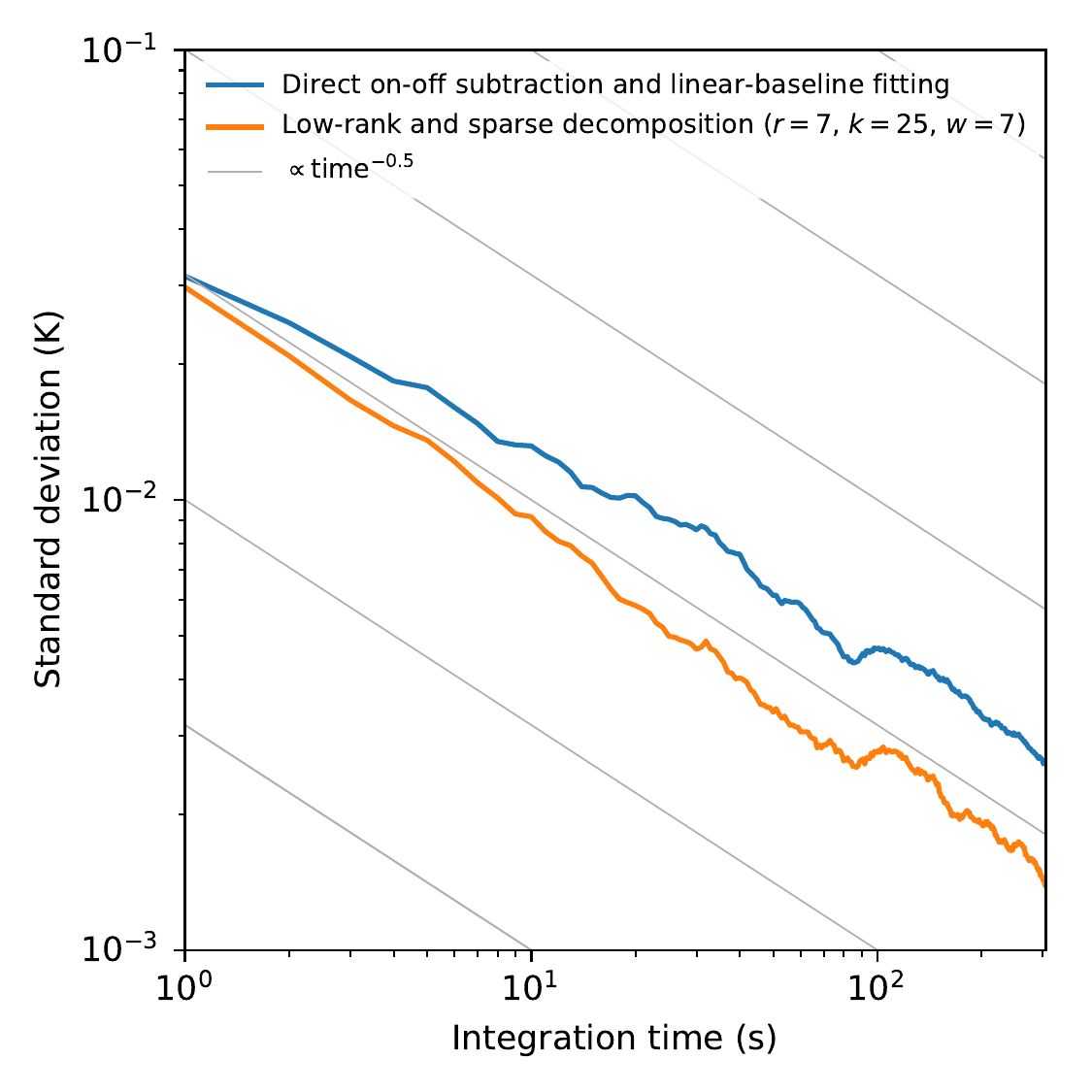}
    \caption{
        Integration time versus achieved standard deviation of emission-free channels in the time-series spectra of the redshifted CO~(5--4) observation reduced by the conventional (blue line) and proposed (orange line) methods.
        To obtain time-series spectra in the former case, we first time-integrate each 10-s off-source measurement and then subtract it from each 1-s on-source measurement.
        Gray sloped lines are proportional to the square root of the integration time (the standard deviation decreases in parallel in the case of white noise).
    }
    \label{fig:time-vs-noise-PJ020941.3-CO54}
\end{figure}

\begin{table*}[h]
    \centering
    \caption{Noise levels of the CO~(5--4) spectra reduced by different methods}
    \label{tab:noise-levels-of-the-CO54-spectra-reduced-by-the-different-methods}
    \begin{tabular}{p{0.26\linewidth}p{0.32\linewidth}p{0.32\linewidth}}
        \hline\hline
        ~ & Achieved standard deviation of emission-free channels & Expected noise level inferred from Equation~\ref{eq:psw-and-cw-noise}\\
        ~ & (mK) & (mK)\\
        \hline
        Conventional method & 2.60 & 2.18\\
        Proposed method & 1.39 & 1.54 ($=2.18/\sqrt{2}$)\\
        \hline
    \end{tabular}
\end{table*}

\clearpage
\section{Glossary lists}
\label{s:glossary-lists}

\begin{table*}[h]
    \centering
    \caption{Glossary of Section~\ref{s:introduction}.}
    \begin{tabular}{p{0.1\linewidth}p{0.2\linewidth}p{0.6\linewidth}}
        \hline\hline
        Notation & First appearance & Description\\
        \hline
        $P$ & Equation~\ref{eq:psw-and-cw} & Output power spectrum of a receiver-spectrometer system.\\
        $T_{\star}$ & Equation~\ref{eq:psw-and-cw} & Brightness temperature of the astronomical signals corrected for the atmospheric transmission. Both signals from a target and cosmic microwave background (CMB) are included. Beam dilution may be corrected to obtain the intrinsic brightness temperature of the target.\\
        $\T{sys}$ & Equation~\ref{eq:psw-and-cw-noise} & System noise temperature.\\
        $\sigma_{\star}$ & Equation~\ref{eq:psw-and-cw-noise} & Standard-deviation noise level of $T_{\star}$.\\
        $\Delta \nu$ & Equation~\ref{eq:psw-and-cw-noise} & Frequency channel width of a spectrometer.\\
        $t\subrm{on}$ & Equation~\ref{eq:psw-and-cw-noise} & Total on-source time of an observation.\\
        \hline
    \end{tabular}
\end{table*}

\begin{table*}[h]
    \centering
    \caption{Glossary of Section~\ref{s:reformulation}.}
    \begin{tabular}{p{0.1\linewidth}p{0.2\linewidth}p{0.6\linewidth}}
        \hline\hline
        Notation & First appearance & Description\\
        \hline
        $k\subrm{B}$ & Equation~\ref{eq:general-response-function} & The Boltzmann constant.\\
        $t$ & Equation~\ref{eq:general-response-function} & Measured time or elapsed time from some origin.\\
        $\nu$ & Equation~\ref{eq:general-response-function} & Observed frequency or radio frequency (RF). Dependent on time, $\nu(t)$, in a frequency modulation observation.\\
        $f$ & Equation~\ref{eq:general-response-function} & Measured frequency in a spectrometer. Equivalent to an intermediate frequency (IF) in a heterodyne receiver. $f=\nu$ for a direct detector.\\
        $m$ & Equation~\ref{eq:frequency-conversion} & Frequency conversion term between the measured and observed frequencies. Here, $|m|$ is equivalent to the local oscillator (LO) frequency in a heterodyne receiver. $m=0$ in a direct detector. Dependent on time, $m(t)$, in a frequency modulation observation.\\
        $\phi$ & Equation~\ref{eq:position-indicator} & Position indicator, which discriminates between on ($\phi(t)=1$) and off-source ($\phi(t)=0$) times during an observation.\\
        $G$ & Equation~\ref{eq:general-response-function} & Gain between an RF input and an IF output.\\
        $\T{in}$ & Equation~\ref{eq:general-response-function} & Input brightness temperature to a receiver. \\
        $\T{noise}$ & Equation~\ref{eq:general-response-function} & Equivalent noise temperature of a receiver-spectrometer system.\\
        $\T{sky}$ & Equation~\ref{eq:input-temperature} & Brightness temperature of the signals from the sky.\\
        $\T{ast}$ & Equation~\ref{eq:unified-sky-temperature} & Brightness temperature of astronomical signals corrected for the atmospheric transmission, which expresses both on-source ($\T{ast}=T_{\star}$) and off-source ($\T{ast}=0$; virtual) measurements.\\
        $\T{amb}$ & Equation~\ref{eq:input-temperature} & Ambient temperature around a telescope.\\
        $\T{room}$ & Equation~\ref{eq:input-temperature} & Room temperature around the receiver.\\
        $\T{atm}$ & Equation~\ref{eq:sky-temperature} & Physical temperature of the atmosphere.\\
        $\eff{fwd}$ & Equation~\ref{eq:input-temperature} & Forward efficiency of telescope feed.\\
        $\eff{atm}$ & Equation~\ref{eq:sky-temperature} & Line-of-sight atmospheric transmission.\\
        $X$ & Equation~\ref{eq:matrix-for-decomposition} & Element of a two-dimensional matrix calculated from the powers of the sky and a hot load and considered to be decomposed into low-rank and sparse matrices.\\
        $L$ & Equation~\ref{eq:lowrank-sparse-decomposition} & Element of a two-dimensional matrix whose rank is expected to be much lower than full rank.\\
        $S$ & Equation~\ref{eq:lowrank-sparse-decomposition} & Element of a two-dimensional matrix whose elements are mostly expected to be zero.\\
        $E$ & Equation~\ref{eq:lowrank-sparse-decomposition} & Noise arose from the atmosphere and a receiver-spectrometer system.\\
        $t\subrm{obs}$ & Equation~\ref{eq:integrated-spectrum} & Total observation time of both on and off-source positions (overheads are not included).\\
        \hline
    \end{tabular}
\end{table*}

\begin{table*}[h]
    \centering
    \caption{Glossary of Section~\ref{s:proposed-method}.}
    \begin{tabular}{p{0.1\linewidth}p{0.2\linewidth}p{0.6\linewidth}}
        \hline\hline
        Notation & First appearance & Description\\
        \hline
        $N\subrm{freq}$ & Equation~\ref{eq:discrete-sample} & The number of frequency sample of a matrix.\\
        $N\subrm{time}$ & Equation~\ref{eq:discrete-sample} & The number of time sample of a matrix.\\
        $k$ & Algorithm~\ref{algo:godec-algorithm} & The number of non-zero elements of a vector or a matrix.\\
        $r$ & Algorithm~\ref{algo:godec-algorithm} & The rank of a low-rank matrix.\\
        SVD & Algorithm~\ref{algo:godec-algorithm} & Singular value decomposition ($\boldsymbol{U}\boldsymbol{\Lambda}\boldsymbol{V}^{T} = \mathrm{SVD}(\boldsymbol{X})$).\\
        $\boldsymbol{\Omega}$ & Algorithm~\ref{algo:godec-algorithm} & Index matrix for sparse identification in a matrix.\\
        $\boldsymbol{\theta}$ & Algorithm~\ref{algo:sparseid-for-fast-sampled-psw-observations} & Index vector for sparse identification in a vector.\\
        $\boldsymbol{s}$ & Algorithm~\ref{algo:sparseid-for-fast-sampled-psw-observations} & Time-integrated on-source spectrum.\\
        $w$ & Figure~\ref{fig:sparseid-for-fast-sampled-psw-observations} & Filter length of a smoothed spectrum.\\
        \hline
    \end{tabular}
\end{table*}

%% file: article.bbl
\begin{thebibliography}{}
\expandafter\ifx\csname natexlab\endcsname\relax\def\natexlab#1{#1}\fi
\providecommand{\url}[1]{\href{#1}{#1}}
\providecommand{\dodoi}[1]{doi:~\href{http://doi.org/#1}{\nolinkurl{#1}}}
\providecommand{\doeprint}[1]{\href{http://ascl.net/#1}{\nolinkurl{http://ascl.net/#1}}}
\providecommand{\doarXiv}[1]{\href{https://arxiv.org/abs/#1}{\nolinkurl{https://arxiv.org/abs/#1}}}

\bibitem[{Cand\`{e}s {et~al.}(2011)Cand\`{e}s, Li, Ma, \& Wright}]{Candes2011}
Cand\`{e}s, E.~J., Li, X., Ma, Y., \& Wright, J. 2011, J. ACM, 58,
  \dodoi{10.1145/1970392.1970395}

\bibitem[{{Cand\`{e}s} {et~al.}(2006){Cand\`{e}s}, {Romberg}, \&
  {Tao}}]{CandesTao2006}
{Cand\`{e}s}, E.~J., {Romberg}, J., \& {Tao}, T. 2006, IEEE Transactions on
  Information Theory, 52, 489, \dodoi{10.1109/TIT.2005.862083}

\bibitem[{Chapin {et~al.}(2013)Chapin, Berry, Gibb, Jenness, Scott, Tilanus,
  Economou, \& Holland}]{Chapin2013}
Chapin, E.~L., Berry, D.~S., Gibb, A.~G., {et~al.} 2013, Monthly Notices of the
  Royal Astronomical Society, 430, 2545, \dodoi{10.1093/mnras/stt052}

\bibitem[{Collaboration {et~al.}(2014)Collaboration, Ade, Aghanim, Arg\"{u}eso,
  Armitage-Caplan, Arnaud, Ashdown, Atrio-Barandela, Aumont, Baccigalupi,
  Banday, Barreiro, Bartlett, Battaner, Beelen, Benabed, Beno\^{\i}t,
  Benoit-L\'{e}vy, Bernard, Bersanelli, Bielewicz, Bobin, Bock, Bonaldi,
  Bonavera, Bond, Borrill, Bouchet, Bridges, Bucher, Burigana, Butler, Cardoso,
  Carvalho, Catalano, Challinor, Chamballu, Chen, Chiang, Chiang, Christensen,
  Church, Clemens, Clements, Colombi, Colombo, Couchot, Coulais, Crill, Curto,
  Cuttaia, Danese, Davies, Davis, Bernardis, Rosa, Zotti, Delabrouille,
  Delouis, D\'{e}sert, Dickinson, Diego, Dole, Donzelli, Dor\'{e}, Douspis,
  Dupac, Efstathiou, En\ss{}lin, Eriksen, Finelli, Forni, Frailis, Franceschi,
  Galeotta, Ganga, Giard, Giardino, Giraud-H\'{e}raud, Gonz\'{a}lez-Nuevo,
  G\'{o}rski, Gratton, Gregorio, Gruppuso, Hansen, Hanson, Harrison,
  Henrot-Versill\'{e}, Hern\'{a}ndez-Monteagudo, Herranz, Hildebrandt, Hivon,
  Hobson, Holmes, Hornstrup, Hovest, Huffenberger, Jaffe, Jaffe, Jones, Juvela,
  Keih\"{a}nen, Keskitalo, Kisner, Kneissl, Knoche, Knox, Kunz, Kurki-Suonio,
  Lagache, L\"{a}hteenm\"{a}ki, Lamarre, Lasenby, Laureijs, Lawrence, Leahy,
  Leonardi, Le\'{o}n-Tavares, Leroy, Lesgourgues, Liguori, Lilje,
  Linden-V\o{}rnle, L\'{o}pez-Caniego, Lubin, Mac\'{\i}as-P\'{e}rez, Maffei,
  Maino, Mandolesi, Maris, Marshall, Martin, Mart\'{\i}nez-Gonz\'{a}lez, Masi,
  Massardi, Matarrese, Matthai, Mazzotta, McGehee, Meinhold, Melchiorri,
  Mendes, Mennella, Migliaccio, Mitra, Miville-Desch\^{e}nes, Moneti, Montier,
  Morgante, Mortlock, Munshi, Murphy, Naselsky, Nati, Natoli, Negrello,
  Netterfield, N\o{}rgaard-Nielsen, Noviello, Novikov, Novikov, O'Dwyer,
  Osborne, Oxborrow, Paci, Pagano, Pajot, Paladini, Paoletti, Partridge,
  Pasian, Patanchon, Pearson, Perdereau, Perotto, Perrotta, Piacentini, Piat,
  Pierpaoli, Pietrobon, Plaszczynski, Pointecouteau, Polenta, Ponthieu, Popa,
  Poutanen, Pratt, Pr\'{e}zeau, Prunet, Puget, Rachen, Reach, Rebolo, Reinecke,
  Remazeilles, Renault, Ricciardi, Riller, Ristorcelli, Rocha, Rosset, Roudier,
  Rowan-Robinson, Rubi\~{n}o Mart\'{\i}n, Rusholme, Sandri, Santos, Savini,
  Scott, Seiffert, Shellard, Spencer, Starck, Stolyarov, Stompor, Sudiwala,
  Sunyaev, Sureau, Sutton, Suur-Uski, Sygnet, Tauber, Tavagnacco, Terenzi,
  Toffolatti, Tomasi, Tristram, Tucci, Tuovinen, T\"{u}rler, Umana, Valenziano,
  Valiviita, Tent, Varis, Vielva, Villa, Vittorio, Wade, Walter, Wandelt, Yvon,
  Zacchei, \& Zonca}]{Planck2014}
Collaboration, P., Ade, P. A.~R., Aghanim, N., {et~al.} 2014, Astronomy \&
  Astrophysics, 571, A28, \dodoi{10.1051/0004-6361/201321524}

\bibitem[{Collaboration: {et~al.}(2013)Collaboration:, Robitaille, Tollerud, \&
  Greenfield}]{Astropy2013}
Collaboration:, T.~A., Robitaille, T.~P., Tollerud, E.~J., \& Greenfield, P.
  2013, Astronomy \& Astrophysics, \dodoi{10.1051/0004-6361/201322068}

\bibitem[{Collaboration {et~al.}(2018)Collaboration, Price-Whelan, Sip\H{o}cz,
  G\"{u}nther, Lim, Crawford, Conseil, Shupe, Craig, Dencheva, Ginsburg,
  VanderPlas, Bradley, P\'{e}rez-Su\'{a}rez, Val-Borro, Aldcroft, Cruz,
  Robitaille, Tollerud, Ardelean, Babej, Bach, Bachetti, Bakanov, Bamford,
  Barentsen, Barmby, Baumbach, Berry, Biscani, Boquien, Bostroem, Bouma,
  Brammer, Bray, Breytenbach, Buddelmeijer, Burke, Calderone, Rodr\'{\i}guez,
  Cara, Cardoso, Cheedella, Copin, Corrales, Crichton, D'Avella, Deil, Depagne,
  Dietrich, Donath, Droettboom, Earl, Erben, Fabbro, Ferreira, Finethy, Fox,
  Garrison, Gibbons, Goldstein, Gommers, Greco, Greenfield, Groener, Grollier,
  Hagen, Hirst, Homeier, Horton, Hosseinzadeh, Hu, Hunkeler, Ivezi\'{c}, Jain,
  Jenness, Kanarek, Kendrew, Kern, Kerzendorf, Khvalko, King, Kirkby, Kulkarni,
  Kumar, Lee, Lenz, Littlefair, Ma, Macleod, Mastropietro, McCully, Montagnac,
  Morris, Mueller, Mumford, Muna, Murphy, Nelson, Nguyen, Ninan, N\"{o}the,
  Ogaz, Oh, Parejko, Parley, Pascual, Patil, Patil, Plunkett, Prochaska,
  Rastogi, Janga, Sabater, Sakurikar, Seifert, Sherbert, Sherwood-Taylor, Shih,
  Sick, Silbiger, Singanamalla, Singer, Sladen, Sooley, Sornarajah, Streicher,
  Teuben, Thomas, Tremblay, Turner, Terr\'{o}n, Kerkwijk, Vega, Watkins,
  Weaver, Whitmore, Woillez, \& Zabalza}]{Astropy2018}
Collaboration, T.~A., Price-Whelan, A.~M., Sip\H{o}cz, B.~M., {et~al.} 2018,
  The Astronomical Journal, 156, 123, \dodoi{10.3847/1538-3881/aabc4f}

\bibitem[{Cort\'{e}s {et~al.}(2016)Cort\'{e}s, Reeves, \& Bustos}]{Cortes2016}
Cort\'{e}s, F., Reeves, R., \& Bustos, R. 2016, Radio Science, 51, 1166,
  \dodoi{10.1002/2015rs005929}

\bibitem[{Dempsey {et~al.}(2013)Dempsey, Friberg, Jenness, Tilanus, Thomas,
  Holland, Bintley, Berry, Chapin, Chrysostomou, Davis, Gibb, Parsons, \&
  Robson}]{Dempsey2013}
Dempsey, J.~T., Friberg, P., Jenness, T., {et~al.} 2013, Monthly Notices of the
  Royal Astronomical Society, 430, 2534, \dodoi{10.1093/mnras/stt090}

\bibitem[{{Donoho}(2006)}]{Donoho2006}
{Donoho}, D.~L. 2006, IEEE Transactions on Information Theory, 52, 1289,
  \dodoi{10.1109/TIT.2006.871582}

\bibitem[{Endo {et~al.}(2019)Endo, Karatsu, Laguna, Mirzaei, Huiting, Thoen,
  Murugesan, Yates, Bueno, Marrewijk, Bosma, Yurduseven, Llombart, Suzuki,
  Naruse, Visser, Werf, Klapwijk, \& Baselmans}]{Endo2019a}
Endo, A., Karatsu, K., Laguna, A.~P., {et~al.} 2019, Journal of Astronomical
  Telescopes, Instruments, and Systems, 5, 1,
  \dodoi{10.1117/1.jatis.5.3.035004}

\bibitem[{{Endo} {et~al.}(2019){Endo}, {Karatsu}, {Tamura}, {Oshima},
  {Taniguchi}, {Takekoshi}, {Asayama}, {Bakx}, {Bosma}, {Bueno}, {Chin},
  {Fujii}, {Fujita}, {Huiting}, {Ikarashi}, {Ishida}, {Ishii}, {Kawabe},
  {Klapwijk}, {Kohno}, {Kouchi}, {Llombart}, {Maekawa}, {Murugesan},
  {Nakatsubo}, {Naruse}, {Ohtawara}, {Pascual Laguna}, {Suzuki}, {Suzuki},
  {Thoen}, {Tsukagoshi}, {Ueda}, {de Visser}, {van der Werf}, {Yates},
  {Yoshimura}, {Yurduseven}, \& {Baselmans}}]{Endo2019b}
{Endo}, A., {Karatsu}, K., {Tamura}, Y., {et~al.} 2019, Nature Astronomy, 3,
  989, \dodoi{10.1038/s41550-019-0850-8}

\bibitem[{Endo {et~al.}(2020)Endo, Karatsu, Tamura, Oshima, Taniguchi,
  Takekoshi, Asayama, Bakx, Bosma, Bueno, Buijtendorp, Chin, Fujii, Fujita,
  Huijten, Huiting, Ikarashi, Ishida, Ishii, Kawabe, Klapwijk, Kohno, Kouchi,
  Llombart, Maekawa, Murugesan, Nakatsubo, Naruse, Ohtawara, Laguna, Suzuki,
  Suzuki, Thoen, Tsukagoshi, Ueda, de~Visser, van~der Werf, Yates, Yoshimura,
  Yurduseven, Dabironezare, Hähnle, \& Baselmans}]{Endo2020}
Endo, A., Karatsu, K., Tamura, Y., {et~al.} 2020, in Millimeter, Submillimeter,
  and Far-Infrared Detectors and Instrumentation for Astronomy X, ed.
  J.~Zmuidzinas \& J.-R. Gao, Vol. 11453, International Society for Optics and
  Photonics (SPIE), \dodoi{10.1117/12.2560669}

\bibitem[{{Erickson} {et~al.}(2007){Erickson}, {Narayanan}, {Goeller}, \&
  {Grosslein}}]{Erickson2007}
{Erickson}, N., {Narayanan}, G., {Goeller}, R., \& {Grosslein}, R. 2007, in
  Astronomical Society of the Pacific Conference Series, Vol. 375, From
  Z-Machines to ALMA: (Sub)Millimeter Spectroscopy of Galaxies, ed. A.~J.
  {Baker}, J.~{Glenn}, A.~I. {Harris}, J.~G. {Mangum}, \& M.~S. {Yun}, 71

\bibitem[{Harrington {et~al.}(2016)Harrington, Yun, Cybulski, Wilson, Aretxaga,
  Chavez, De~la Luz, Erickson, Ferrusca, Gallup, Hughes, Monta\~{n}a,
  Narayanan, S\'{a}nchez-Arg\"{u}elles, Schloerb, Souccar, Terlevich,
  Terlevich, Zeballos, \& Zavala}]{Harrington2016}
Harrington, K.~C., Yun, M.~S., Cybulski, R., {et~al.} 2016, Monthly Notices of
  the Royal Astronomical Society, 458, 4383, \dodoi{10.1093/mnras/stw614}

\bibitem[{Harris {et~al.}(2020)Harris, Millman, van~der Walt, Gommers,
  Virtanen, Cournapeau, Wieser, Taylor, Berg, Smith, Kern, Picus, Hoyer, van
  Kerkwijk, Brett, Haldane, del R{'{\i}}o, Wiebe, Peterson,
  G{'{e}}rard-Marchant, Sheppard, Reddy, Weckesser, Abbasi, Gohlke, \&
  Oliphant}]{Harris2020}
Harris, C.~R., Millman, K.~J., van~der Walt, S.~J., {et~al.} 2020, Nature, 585,
  357, \dodoi{10.1038/s41586-020-2649-2}

\bibitem[{Heiles(2007)}]{Heiles2007}
Heiles, C. 2007, Publications of the Astronomical Society of the Pacific, 119,
  643, \dodoi{10.1086/519532}

\bibitem[{Hoyer \& Hamman(2016)}]{Hoyer2017}
Hoyer, S., \& Hamman, J. 2016, Journal of Open Research Software, 5,
  \dodoi{10.5334/jors.148}

\bibitem[{Hunter(2007)}]{Hunter2007}
Hunter, J.~D. 2007, Computing in Science \& Engineering, 9, 90,
  \dodoi{10.1109/MCSE.2007.55}

\bibitem[{{Iwai} {et~al.}(2017){Iwai}, {Kubo}, {Ishibashi}, {Naoi}, {Harada},
  {Ema}, {Hayashi}, \& {Chikahiro}}]{Iwai2017}
{Iwai}, K., {Kubo}, Y., {Ishibashi}, H., {et~al.} 2017, Earth, Planets, and
  Space, 69, 95, \dodoi{10.1186/s40623-017-0681-8}

\bibitem[{Kawabe {et~al.}(2016)Kawabe, Kohno, Tamura, Takekoshi, Oshima, \&
  Ishii}]{Kawabe2016}
Kawabe, R., Kohno, K., Tamura, Y., {et~al.} 2016, in Ground-based and Airborne
  Telescopes VI, ed. H.~J. Hall, R.~Gilmozzi, \& H.~K. Marshall, Vol. 9906,
  International Society for Optics and Photonics (SPIE), 779 -- 790,
  \dodoi{10.1117/12.2232202}

\bibitem[{{Klaassen} {et~al.}(2019){Klaassen}, {Mroczkowski}, {Bryan},
  {Groppi}, {Basu}, {Cicone}, {Dannerbauer}, {De Breuck}, {Fischer}, {Geach},
  {Hatziminaoglou}, {Holland}, {Kawabe}, {Sehgal}, {Stanke}, \& {van
  Kampen}}]{Klaassen2019}
{Klaassen}, P., {Mroczkowski}, T., {Bryan}, S., {et~al.} 2019, in Bulletin of
  the American Astronomical Society, Vol.~51, 58.
\newblock \doarXiv{1907.04756}

\bibitem[{Klein {et~al.}(2012)Klein, Hochg\"{u}rtel, Kr\"{a}mer, Bell, Meyer,
  \& G\"{u}sten}]{Klein2012}
Klein, B., Hochg\"{u}rtel, S., Kr\"{a}mer, I., {et~al.} 2012, Astronomy \&
  Astrophysics, 542, L3, \dodoi{10.1051/0004-6361/201218864}

\bibitem[{{Kohno} {et~al.}(2019){Kohno}, {Tamura}, {Inoue}, {Kawabe}, {Oshima},
  {Hatsukade}, {Takekoshi}, {Yoshimura}, {Umehata}, {Dannerbauer}, {Cicone}, \&
  {Bertoldi}}]{Kohno2019}
{Kohno}, K., {Tamura}, Y., {Inoue}, A., {et~al.} 2019, Astro2020: Decadal
  Survey on Astronomy and Astrophysics, 2020, 402.
\newblock \doarXiv{1907.03488}

\bibitem[{Kojima {et~al.}(2020)Kojima, Kiuchi, Uemizu, Uzawa, Kroug, Gonzalez,
  Dippon, \& Kageura}]{Kojima2020}
Kojima, T., Kiuchi, H., Uemizu, K., {et~al.} 2020, Astronomy \& Astrophysics,
  640, L9, \dodoi{10.1051/0004-6361/202038713}

\bibitem[{Lou {et~al.}(2020)Lou, xi~Zuo, jun Yao, cai Shi, Yang, \& peng
  Chen}]{Lou2020}
Lou, Z., xi~Zuo, Y., jun Yao, Q., {et~al.} 2020, Appl. Opt., 59, 3353,
  \dodoi{10.1364/AO.388320}

\bibitem[{McKinney(2010)}]{McKinney2010}
McKinney, W. 2010, in Proceedings of the 9th Python in Science Conference, Vol.
  445, Austin, TX, 51--56

\bibitem[{McKinney(2011)}]{McKinney2011}
McKinney, W. 2011, Python for High Performance and Scientific Computing, 14

\bibitem[{Morii {et~al.}(2017)Morii, Ikeda, Sako, \& Ohsawa}]{Morii2017}
Morii, M., Ikeda, S., Sako, S., \& Ohsawa, R. 2017, The Astrophysical Journal,
  835, 1, \dodoi{10.3847/1538-4357/835/1/1}

\bibitem[{Pedregosa {et~al.}(2011)Pedregosa, Varoquaux, Gramfort, Michel,
  Thirion, Grisel, Blondel, Prettenhofer, Weiss, Dubourg, Vanderplas, Passos,
  Cournapeau, Brucher, Perrot, \& Duchesnay}]{Pedregosa2011}
Pedregosa, F., Varoquaux, G., Gramfort, A., {et~al.} 2011, Journal of Machine
  Learning Research, 12, 2825

\bibitem[{Schieder \& Kramer(2001)}]{Schieder2001}
Schieder, R., \& Kramer, C. 2001, Astronomy \& Astrophysics, 373, 746,
  \dodoi{10.1051/0004-6361:20010611}

\bibitem[{Schloerb(2008)}]{Schloerb2008}
Schloerb, F.~P. 2008, in Ground-based and Airborne Telescopes II, ed. L.~M.
  Stepp \& R.~Gilmozzi, Vol. 7012, International Society for Optics and
  Photonics (SPIE), 299 -- 310, \dodoi{10.1117/12.791296}

\bibitem[{Taniguchi {et~al.}(2019)Taniguchi, Tamura, Kohno, Takahashi,
  Horigome, Maekawa, Sakai, Kuno, \& Minamidani}]{Taniguchi2020}
Taniguchi, A., Tamura, Y., Kohno, K., {et~al.} 2019, Publications of the
  Astronomical Society of Japan, 72, \dodoi{10.1093/pasj/psz121}

\bibitem[{{The EHT Collaboration, et al.}(2019)}]{EHTcollaborationIV}
{The EHT Collaboration, et al.} 2019, ApJL, 875, 4,
  \dodoi{10.3847/2041-8213/ab0e85}

\bibitem[{Tibshirani(1996)}]{Tibshirani1996}
Tibshirani, R. 1996, Journal of the Royal Statistical Society. Series B
  (Methodological), 58, 267, \dodoi{10.1111/j.2517-6161.1996.tb02080.x}

\bibitem[{Uemura {et~al.}(2015)Uemura, Kawabata, Ikeda, \& Maeda}]{Uemura2015}
Uemura, M., Kawabata, K.~S., Ikeda, S., \& Maeda, K. 2015, Publications of the
  Astronomical Society of Japan, 67, \dodoi{10.1093/pasj/psv031}

\bibitem[{{Ungerechts} {et~al.}(2000){Ungerechts}, {Brunswig}, {Penalver},
  {Perrigouard}, \& {Sievers}}]{Ungerechts2000}
{Ungerechts}, H., {Brunswig}, W., {Penalver}, J., {Perrigouard}, A., \&
  {Sievers}, A. 2000, in Society of Photo-Optical Instrumentation Engineers
  (SPIE) Conference Series, Vol. 4009, Advanced Telescope and Instrumentation
  Control Software, ed. H.~{Lewis}, 327--337, \dodoi{10.1117/12.388403}

\bibitem[{Viero {et~al.}(2014)Viero, Asboth, Roseboom, Moncelsi, Marsden,
  Cooper, Zemcov, Addison, Baker, Beelen, Bock, Bridge, Conley, Devlin,
  Dor\'{e}, Farrah, Finkelstein, Font-Ribera, Geach, Gebhardt, Gill, Glenn,
  Hajian, Halpern, Jogee, Kurczynski, Lapi, Negrello, Oliver, Papovich, Quadri,
  Ross, Scott, Schulz, Somerville, Spergel, Vieira, Wang, \&
  Wechsler}]{Viero2014}
Viero, M.~P., Asboth, V., Roseboom, I.~G., {et~al.} 2014, The Astrophysical
  Journal Supplement Series, 210, 22, \dodoi{10.1088/0067-0049/210/2/22}

\bibitem[{{Virtanen} {et~al.}(2020){Virtanen}, {Gommers}, {Oliphant},
  {Haberland}, {Reddy}, {Cournapeau}, {Burovski}, {Peterson}, {Weckesser},
  {Bright}, {van der Walt}, {Brett}, {Wilson}, {Jarrod Millman}, {Mayorov},
  {Nelson}, {Jones}, {Kern}, {Larson}, {Carey}, {Polat}, {Feng}, {Moore}, {Vand
  erPlas}, {Laxalde}, {Perktold}, {Cimrman}, {Henriksen}, {Quintero}, {Harris},
  {Archibald}, {Ribeiro}, {Pedregosa}, {van Mulbregt}, \&
  {Contributors}}]{Virtanen2020}
{Virtanen}, P., {Gommers}, R., {Oliphant}, T.~E., {et~al.} 2020, Nature
  Methods, 17, 261, \dodoi{10.1038/s41592-019-0686-2}

\bibitem[{Wheeler {et~al.}(2016)Wheeler, Hailey-Dunsheath, Shirokoff, Barry,
  Bradford, Chapman, Che, Glenn, Hollister, Kov\'{a}cs, LeDuc, Mauskopf,
  McGeehan, McKenney, O'Brient, Padin, Reck, Ross, Shiu, Tucker, Williamson, \&
  Zmuidzinas}]{Wheeler2016}
Wheeler, J., Hailey-Dunsheath, S., Shirokoff, E., {et~al.} 2016, Millimeter,
  Submillimeter, and Far-Infrared Detectors and Instrumentation for Astronomy
  VIII, 99143K, \dodoi{10.1117/12.2233798}

\bibitem[{Wilson {et~al.}(2012)Wilson, Rohlfs, \& Huttemeister}]{Wilson2012}
Wilson, T.~L., Rohlfs, K., \& Huttemeister, S. 2012, Tools of Radio Astronomy,
  \dodoi{10.1007/978-3-540-85122-6}

\bibitem[{Zhou \& Tao(2011)}]{Zhou2011}
Zhou, T., \& Tao, D. 2011, in Proceedings of the 28th International Conference
  on Machine Learning (ICML-11), ed. L.~Getoor \& T.~Scheffer, ICML '11 (New
  York, NY, USA: ACM), 33--40, \dodoi{10.5555/3104482.3104487}

\bibitem[{Zuo {et~al.}(2018)Zuo, Chen, Ansari, \& Lu}]{Zuo2018}
Zuo, S., Chen, X., Ansari, R., \& Lu, Y. 2018, The Astronomical Journal, 157,
  4, \dodoi{10.3847/1538-3881/aaef3b}

\end{thebibliography}
